# The Image of the Process Interpretation of Regular Expressions Is Not Closed under Bisimulation Collapse

Clemens Grabmayer
*Computer Science Department*
*Gran Sasso Science Institute*
*67100 L'Aquila AQ, Italy*
*Email:* `clemens.grabmayer@gssi.it`

*Abstract*—Axiomatization and expressibility problems for Milner's process semantics (1984) of regular expressions modulo bisimilarity have turned out to be difficult for the full class of expressions with deadlock 0 and empty step 1. We report on a phenomenon that arises from the added presence of 1 when 0 is available, and that brings a crucial reason for this difficulty into focus. To wit, while the set of interpretations of 1-free regular expressions is closed under bisimulation collapse, this is not the case for the set of interpretations of all regular expressions.

Process graph interpretations of 1-free regular expressions satisfy the loop existence and elimination property LEE, which is preserved under bisimulation collapse. These features of LEE were applied for showing that an equational proof system for 1-free regular expressions modulo bisimilarity is complete, and that it is decidable in polynomial time whether a process graph is bisimilar to the interpretation of a 1-free regular expression.

While interpretations of regular expressions do not satisfy the property LEE in general, we show that LEE can be recovered by refined interpretations as graphs with 1-transitions (which are similar to silent steps for automata). This suggests that LEE can be expedient also for the general axiomatization and expressibility problems. But a new phenomenon emerges that needs to be addressed: the property of a process graph 'to can be refined into a process graph with 1-transitions and with LEE' is not preserved under bisimulation collapse. We provide a 10-vertex graph with two 1-transitions that satisfies LEE, and in which a pair of bisimilar vertices cannot be collapsed on to each other while preserving the refinement property. This implies that the image of the process interpretation of regular expressions is not closed under bisimulation collapse.

## 1. Introduction

Milner [1] (1984) introduced an interpretation of regular expressions as processes: the interpretation of $0$ is deadlock, of $1$ is successful termination, letters $a$ are atomic actions, the operators $+$ and $\cdot$ stand for choice and concatenation of processes, and (unary) Kleene star $(\cdot)^*$ represents iteration with the option to terminate successfully before and after each pass-through. On the basis of this interpretation, Milner was interested in the process semantics that arises by mapping a star expression to its 'star behavior': the bisimilarity equivalence class (the 'behavior') of its process interpretation. He used the term 'star expressions' for regular expressions when they are interpreted as processes.

We will be concerned with a finer analysis of process interpretations of star expressions (finite process graphs) that are bisimilar, and which therefore are contained in the same star behavior (the appertaining bisimilarity equivalence class). Specifically, we will study whether the bisimulation collapse of the process interpretation of a star expression can always be construed as the process interpretation of some (possibly different) star expression. (Note that the process interpretation of a star expression and its bisimulation collapse are bisimilar, and belong to the same star behavior.)

In the context of automata and language theory, an interpretation that matches Milner's process interpretation was described by Antimirov [2] (1996) via 'partial derivatives' of regular expressions. He used this concept to define, for every regular expression, a non-deterministic finite-state automaton (NFA) which is typically much smaller than the deterministic FA (DFA) that is obtained by the standard translation. Although Antimirov viewed automata only as language acceptors (but not as processes), the arising NFAs correspond directly to the process graphs that are specified by a slight variation (see Def. 2.4) of Milner's interpretation.

Unlike for the standard language semantics, where every language accepted by a finite-state automaton is the interpretation of some regular expression, there are finite process graphs that are not bisimilar to the process interpretation of any star expression. This holds for the process graphs $G_1$ and $G_2$ below (as shown by Bosscher [3], and Milner [1]). But the process graph $G_3$ is the interpretation of a star expression (see Ex. 2.7).

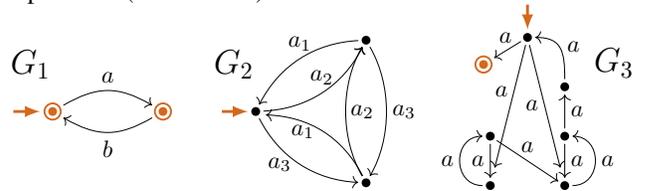

In pictures of process graphs we highlight, here and henceforth, the *start vertex* by a brown arrow ➤ , and a vertex $v$ *with immediate termination* by emphasizing $v$ in brown as ⊙ including a boldface ring.

That the image of the process semantics, even when extended modulo bisimilarity, does not cover all finite process graphs, led Milner to the first of the following questions [1], where the second one is concerned with algebraic properties and axiomatizability of star expressions modulo bisimilarity: **(E)** How can one characterize the process graphs that are expressible by star expressions, that is, bisimilar to ones in the image of the process semantics? **(A)** Is a natural adaptation of Salomaa's complete proof system [4] for language equivalence of regular expressions complete for bisimilarity of the process interpretation of star expressions?

While the decision problem underlying **(E)** has been shown to be solvable [5] (but only with a super-exponential complexity bound), so far only partial solutions have been obtained for question **(A)**. These concern tailored restrictions of Milner's proof system that were shown to be complete for the following subclasses of star expressions: (a) without 0 and 1, but with binary star iteration $e_1 \circledast e_2$ with iteration-part $e_1$ and exit-part $e_2$ instead of unary star [6], (b) with 0, and with iterations restricted to exit-less ones $(\cdot)^* \cdot 0$ in absence of 1 [7] and in the presence of 1 [8], (c) without 0, and with only restricted occurrences of 1 [9], and (d) '1-free' expressions formed with 0, without 1, but again with binary instead of unary iteration [10]. While the classes (c) and (d) are incomparable, these results can be joined to apply to an encompassing proper subclass of the star expressions [10]. These partial results for **(A)** also yield partial results concerning **(E)**: expressibility modulo bisimilarity of a finite process graph by the process interpretation of a star expression in one of these classes is decidable in polynomial time.

**The purpose of this paper.** We describe a phenomenon that can help to explain a subjective experience: that trying to solve the problems **(E)** and **(A)** for the full class of star expressions is much harder than in [10] for the subclass of 1-free star expressions where 0 is present, but 1 is not. It provides evidence that minimization strategies (see below) for solving **(E)** and **(A)** face a significant obstacle in general. This notwithstanding we also report on the investigation by which we made this discovery, and by the continuation of which we expect progress towards solving the problems.

**Minimization strategies.** All approaches to the problems **(E)** and **(A)** we are aware of use a minimization strategy for expressions (most frequently), or for the associated process graphs or specifications of these graphs (less often). For the axiomatization problem **(A)**, the approach on expression level proceeds as follows. In order to prove equal, in a proof system $\mathcal{S}$, two terms $e_1$ and $e_2$ with bisimilar interpretations, the strategy aims to simplify $e_1$ and $e_2$ as much as possible by algebraic operations that preserve the interpretations up to bisimilarity. Let the results be expressions $f_1$ and $f_2$. Ideally, $f_1$ and $f_2$ coincide if simplification is optimal and confluent, then yielding $f_1 = f_2$ directly. Or otherwise, if the reached expressions are simple enough, they can facilitate a proof $f_1 = f_2$ in $\mathcal{S}$ by structural induction. From this proof a derivation of $e_1 = e_2$ in $\mathcal{S}$ is obtained by justifying the simplification steps as proofs of $e_1 = f_1$ and $e_2 = f_2$ in $\mathcal{S}$, and by applying the transitivity rule of equational logic.

A minimization strategy was also used in [11] for showing that the decision problem underlying **(E)** is solvable. Here minimization acts on 'well-behaved' recursive specifications that specify star expressions. Such specifications are simplified via rewrite rules to the set of their normal forms, which can be described formally. The rewrite rules are, however, not confluent, and so neither is simplification. But the problem of checking whether a finite process graph $G$ is expressible can be reduced to the computable problem of checking whether $G$ is bisimilar to a well-behaved specification from a finite set that is computable from $G$.

Another minimization strategy for structure-constrained graphs that correspond to 1-free star expressions with 0 and binary iteration was introduced in [10]. We describe it below.

**Structure-constrained process graphs.** Process interpretations $[\![e]\!]_P$ of 1-free star expressions $e$ possess a structural property called 'Loop Existence and Elimination' (LEE), see below. Vice versa, finite process graphs with LEE are rather directly expressible by star expressions modulo bisimilarity. This was observed in [10], where LEE was introduced, based on the concept of 'loop subgraph'.

A *loop subgraph* of a process graph is generated from a set $E$ of entry transitions from a vertex $v$ by all paths from $v$ that start along a transition in $E$ and continue until $v$ is reached again first, given that three properties hold for the so-constructed subgraph: (L1) there is an infinite path from $v$ starting with an $E$-transition, (L2) every infinite path starting from $v$ with an $E$-transition returns to $v$ eventually, and (L3) termination is permitted only at $v$ (but not required). For example, neither $G_1$ nor $G_2$ on page 1 are loop graphs, that is, loop subgraphs of themselves with $E$ the set of all transitions from the start vertex: $G_1$ violates (L3), and $G_2$ violates (L2), because it facilitates infinite paths that do not return to the start vertex. Moreover, neither $G_1$ and $G_2$ contain loop subgraphs. But the graph $G_3$ has loop subgraphs, see below.

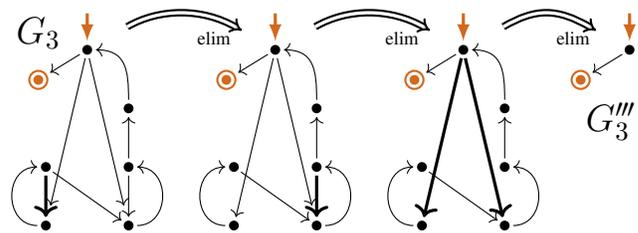

A graph satisfies *Loop Existence and Elimination (LEE)* if repeatedly picking a loop subgraph, eliminating its entry transitions, and performing garbage collection, leads to a graph without infinite paths. The graphs $G_1$ and $G_2$ do not satisfy LEE, because neither contains a loop subgraph that can be eliminated, yet both facilitate infinite paths. But the graph $G_3$ satisfies LEE. The picture below shows a run of the loop elimination procedure for the graph $G_3$. The loop-entry transitions of loop subcharts that are eliminated are marked in bold. We have neglected action labels here. Since the graph $G_3''''$ that is reached after three loop-subgraph elimination steps from $G_3$ does not have an infinite path, we conclude that $G_3$ satisfies LEE.

Expressions correspond to graphs with LEE, see [10]:

**(I)₁** $\llbracket e \rrbracket_P$ satisfies LEE, for every 1-free star expressions $e$.

**Minimization of structure-constrained graphs.** The completeness proof of a tailoring BBP (due Bergstra, Bethke, and Ponse [12]) of Milner's proof system for 1-free star expressions in LEE can be based on **(I)₁** and the following minimization statements:

**(C)** LEE is preserved under bisimulation collapse.

**(IC)₁** The image under $\llbracket \cdot \rrbracket_P$ of the 1-free star expressions is closed under bisimulation collapse. (See Prop. 2.12.)

That property **(IC)₁** holds, hinges on a variation of the definition of the process semantics used in [10], see Def. 2.4.

The completeness proof for BBP proceeds, roughly, as follows. Suppose that 1-free star expressions $e_1$ and $e_2$ have bisimilar process interpretations, formally, $\llbracket e_1 \rrbracket_P \leftrightarrow \llbracket e_2 \rrbracket_P$. We have to obtain a derivation of $e_1 = e_2$ in BBP. As $e_1$ and $e_2$ are bisimilar, they have a joint bisimulation collapse $G_0$, such that $\llbracket e_1 \rrbracket_P \Rightarrow G_0 \Leftarrow \llbracket e_2 \rrbracket_P$ holds, where by $\Rightarrow$ we indicate functional bisimilarity. Since $\llbracket e_1 \rrbracket_P$ and $\llbracket e_2 \rrbracket_P$ satisfy LEE by **(I)₁**, it follows by **(IC)₁** that there must exist a 1-free star expression $e_0$ such that $\llbracket e_0 \rrbracket_P = G_0$. By utilizing the functional bisimulations from $\llbracket e_1 \rrbracket_P$ and $\llbracket e_2 \rrbracket_P$ to $G_0$ it is possible to construct derivations of $e_0 = e_1$ and $e_0 = e_2$ in BBP. These derivations can be combined via transitivity to a derivation of $e_1 = e_2$ in BBP.

**Generalization to the full class of star expressions?** When trying to extend the minimization-strategy approach, either on the expression level or on the graph level, to one that can handle the full class of star expressions, one encounters a complicated picture of cases that have to be dealt with. Although a subjective impression, it pointed to an increased level of difficulty that might have a conceptual explanation.

Hopes that the minimization approach of structure-constrained process graphs as used for 1-free star expressions can be extended to the full class were raised by a two-part result in [13] about a failure and a remedy: On the one hand, LEE does not hold for process interpretations of star expression in general, so the generalization **(I)₁** of **(I)₁** fails:

**(I)₁** $\llbracket e \rrbracket_P$ does *not satisfy* LEE in general. (See Ex. 2.8)

But on the other hand, LEE can be recovered by defining bisimilar variants in the form of process graphs with '1-transitions' that represent empty-step processes.

**Our contribution: structure-constrained minimization encounters a limit.** Sharpening the remedy in [13] for the failure of **(I)₁**, we restore a weakened version **(RI)₁** by showing that the process semantics can be refined into a variant that ensures LEE. Then we prove a crucial difference with maximal minimization for 1-free star expressions, namely, that the collapse and image-closedness statements **(C)** and **(IC)₁** do not generalize in a similar way. In summary:

**(RI)₁** $\llbracket e \rrbracket_P$ can be refined into a process graph with 1-transitions that satisfies LEE, for every star expression $e$. (See Thm. 4.7. A similar result is described in [13].)

**(RC)** The property 'can be refined into a process graph with 1-transitions and LEE' is *not preserved* under bisimulation collapse. (See counterexample in Thm. 5.7.)

**(IC)₁** The image under $\llbracket \cdot \rrbracket_P$ of the star expressions is *not closed* under bisimulation collapse. (See Thm. 5.8.)

We witness the failure of both **(RC)** and **(IC)₁** by a concrete counterexample that derives from a 1-chart that refines the process interpretation of a star expression (see Ex. 5.1).

**Outlook: structure-constrained collapse approximations.** We informally explain that the counterexample to **(RC)** and **(IC)** arises from a specific case in which bisimilar vertices in a 1-chart with LEE cannot be collapsed on to each other while preserving LEE. We found this case in a systematic investigation of when collapse operations are possible LEE-preservingly. That opens up a promising new approach for the general problems **(A)** and **(E)**. It can lead to a procedure for obtaining, for every finite process graph with 1-transitions and LEE, an approximation of its bisimulation collapse that satisfies LEE and in which every vertex has at most one bisimilar counterpart.

**Overview.** In Sect. 2 we gather basic definitions of star expressions and of their process interpretations as 'charts' (finite process graphs). We note that **(IC)₁** can be added to the results for 1-free star expressions in [10], and that **(I)₁** fails in general. As a remedy we introduce '1-charts' in Sect. 3 as charts with separate 1-transitions that represent empty step processes, and we define a 1-chart interpretation of star expressions that refines the chart interpretation. In Sect. 4 we define 'layered LEE-witnesses' (LLEE-witnesses), and show that every 1-chart interpretation of a star expression has a LLEE-witness, yielding **(RI)₁**.

In Sect. 5 we give a 1-chart that yields, together with its collapses, a counterexample to **(RC)** and **(IC)**. We also explain the origin of that example. Finally in Sect. 6 we summarize our results, and report on a promising approach for circumventing the phenomenon that we describe here.

*Please find in the appendix details of proofs that have been omitted or are only sketched.*

## 2. Process semantics of star expressions

**Definition 2.1.** Let $A$ be a set whose members we call *actions*. The set $StExp(A)$ of *star expressions over actions in $A$* is defined by the following grammar, where $a \in A$:

$$e, e_1, e_2 ::= 0 \mid 1 \mid a \mid e_1 + e_2 \mid e_1 \cdot e_2 \mid e^*$$

We call a star expression *normed* (and respectively, 1-*free*, and *under-star-1-free*) if it can be generated for nonterminal $n$ (resp., for nonterminal $f$, and for the nonterminal $uf$) in the more specific grammars below, where $a$ stands for any action in $A$, and $e$ for any star expression in $StExp(A)$:

$$n, n_1, n_2 ::= 1 \mid a \mid n + e \mid e + n \mid n_1 \cdot n_2 \mid n^*$$
$$f, f_1, f_2 ::= 0 \mid a \mid f_1 + f_2 \mid f_1 \cdot f_2 \mid f_1^* \cdot f_2$$
$$uf, uf_1, uf_2 ::= 0 \mid 1 \mid a \mid uf_1 + uf_2 \mid uf_1 \cdot uf_2 \mid f^*$$

By $StExp^{(1)}(A)$ (by $StExp^{(*/1)}(A)$) we denote the set of 1-free (resp., of under-star-1-free) star expressions over $A$.

The *(syntactic) star height* $|e|_*$ of a star expression $e \in StExp(A)$ is the maximal nesting depth of stars in $e$, defined inductively by: $|0|_* := |1|_* := |a|_* := 0$, $|e_1 + e_2|_* := |e_1 \cdot e_2|_* := \max\{|e_1|_*, |e_2|_*\}$, and $|e^*|_* := 1 + |e|_*$.

**Remark 2.2.** In Def. 2.4 and Rem. 2.6 we will need normed star expressions. 1-Free star expressions arise as translations of '1-free star expressions' as defined [10] with 0, without 1, but with binary star iteration $^\circledast$ instead of unary iteration as we use here: thereby binary iterations $f_1{}^\circledast f_2$ are translated to $f_1^* \cdot f_2$. Under-star-1-free star expressions extend the 1-free star expressions with unary star by permitting 1 outside of iterations (see Lem. 2.11), and permitting that outermost iterations may have an empty exit part. Under-star-1-free star expressions are closed under taking derivatives (see Def. 2.4), see Lem. 2.11.

**Definition 2.3.** A *chart* is a 5-tuple $\mathcal{C} = \langle V, A, v_s, \rightarrow, \downarrow \rangle$ that consists of a set $V$ of *vertices*, a set $A$ of *action labels*, the *start vertex* $v_s \in V$ (hence $V \neq \varnothing$), the *labeled transition relation* $\rightarrow \subseteq V \times A \times V$, and the set $\downarrow \subseteq V$ of *vertices with immediate termination*, for short, the *terminating vertices*. We write $s_1 \xrightarrow{a} s_2$ for a transition $\langle s_1, a, s_2 \rangle \in \rightarrow$, and $s\downarrow$ for a terminating state $s \in \downarrow$.

**Definition 2.4.** The *chart interpretation* of a star expression $e \in StExp(A)$ is the chart:

$$\mathcal{C}(e) = \langle V(e), A, e, \rightarrow \cap V(e) \times A \times V(e), \downarrow \cap V(e) \rangle,$$

where $V(e)$ consists of all star expressions that are reachable from $e$ via transitions of the labeled transition relation $\rightarrow \subseteq StExp(A) \times A \times StExp(A)$ that is defined, together with the immediate-termination relation $\downarrow \subseteq StExp(A)$, via derivability in the transition system specification (TSS) $\mathcal{T}(A)$, where $a \in A$, $e, e_1, e_2, e_i, e_i', e' \in StExp(A)$ with $i \in \{1, 2\}$:

$$\frac{}{1\downarrow} \quad \frac{e_i\downarrow}{(e_1+e_2)\downarrow} \quad \frac{e_1\downarrow \quad e_2\downarrow}{(e_1 \cdot e_2)\downarrow} \quad \frac{}{(e^*)\downarrow}$$

$$\frac{}{a \xrightarrow{a} 1} \quad \frac{e_i \xrightarrow{a} e_i'}{e_1 + e_2 \xrightarrow{a} e_i'} \quad \frac{e_1\downarrow \quad e_2 \xrightarrow{a} e_2'}{e_1 \cdot e_2 \xrightarrow{a} e_2'}$$

$$\frac{e_1 \xrightarrow{a} e_1'}{e_1 \cdot e_2 \xrightarrow{a} e_1' \cdot e_2} \text{ (if } e_1' \text{ is normed)} \quad \frac{e_1 \xrightarrow{a} e_1'}{e_1 \cdot e_2 \xrightarrow{a} e_1'} \text{ (if } e_1' \text{ is not normed)}$$

$$\frac{e \xrightarrow{a} e'}{e^* \xrightarrow{a} e' \cdot e^*} \text{ (if } e' \text{ is normed)} \quad \frac{e \xrightarrow{a} e'}{e^* \xrightarrow{a} e'} \text{ (if } e' \text{ is not normed)}$$

If $e \xrightarrow{a} e'$ is derivable in $\mathcal{T}(A)$, for $e, e' \in StExp(A)$, $a \in A$, then we say that $e'$ is a *derivative* of $e$. If $e\downarrow$ is derivable in $\mathcal{T}(A)$, then we say that $e$ *permits immediate termination*.

**Lemma 2.5.** *For every star expression $e \in StExp(A)$, the chart interpretation $\mathcal{C}(e)$ is a finite chart.*

**Remark 2.6.** The TSS $\mathcal{T}(A)$ in Def. 2.4 refines a simpler TSS (see [11]), which we here call $\mathcal{T}_0(A)$, that arises from $\mathcal{T}(A)$ by dropping the two rules with side-condition 'if $e'$ is not normed', and dropping the side-condition 'if $e_1'$ is normed' from the two other rules closeby. Note that an expression is normed if and only if it enables a sequence of transitions to an expression with immediate termination. Without the requirement of normedness in $\mathcal{T}(A)$, the chart interpretation defined for $\mathcal{T}_0(A)$ is not closed under bisimulation collapse for a trivial reason: from a star expression different derivatives $e \cdot f_1$ and $e \cdot f_2$ with $e$ not normed can arise, whose behavior is the same (namely that of $e$). For example, $a \cdot 0 + (a \cdot 0) \cdot a$ would have the derivatives $1 \cdot 0$ and $(1 \cdot 0) \cdot a$, which do not permit transitions nor immediate termination, whereas in $\mathcal{T}(A)$ this expression only has $1 \cdot 0$ as a derivative. For $\mathcal{C}(\cdot)$ as defined above via $\mathcal{T}(A)$ the question of closedness of the image of $\mathcal{C}(\cdot)$ under bisimulation collapse is more challenging, see Prop. 2.12, and Thm. 5.8.

**Example 2.7.** The graph $G_3$ on page 1 is the process interpretation of: $(1 \cdot ((a \cdot a \cdot (a \cdot a)^* \cdot a \cdot a + a) \cdot a \cdot (a \cdot a)^* \cdot a \cdot a)^*) \cdot a$, where we assume associativity to the left for dropped brackets.

**Example 2.8.** For the star expression $e = (a^* \cdot b^*)^*$, the chart interpretation $\mathcal{C}(e)$ as obtained by the TSS $\mathcal{T}(A)$ is:

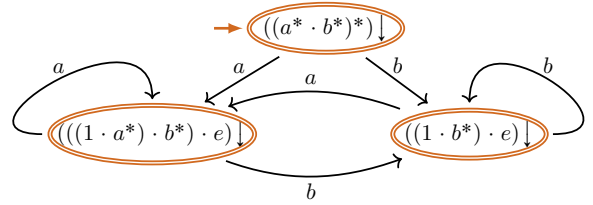

Note that all three vertices permit immediate termination.

For the chart $\mathcal{C}(e)$ we find that it does not satisfy LEE. The following run of the loop elimination procedure:

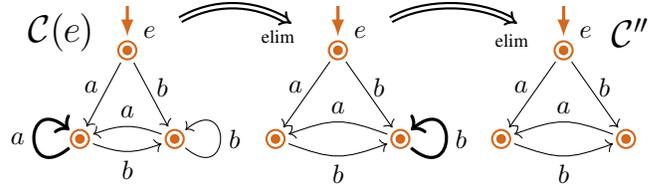

yields a chart $\mathcal{C}''$ that does not contain a loop subchart any more, but still admits infinite paths. Therefore $e$ witnesses (I)$_\mathrm{T}$ in Sect. 1, that process interpretations of regular expressions do not always satisfy the structural property LEE.

**Definition 2.9.** Let $\mathcal{C}_i = \langle V_i, A, v_{s,i}, \rightarrow_i, \downarrow_i \rangle$ be charts, for $i \in \{1, 2\}$, with joint action set $A$. By a *bisimulation between $\mathcal{C}_1$ and $\mathcal{C}_2$* we mean a binary relation $B \subseteq V_1 \times V_2$ such that:

(start) $\langle v_{s,1}, v_{s,2} \rangle \in B$

holds, and for every $\langle v_1, v_2 \rangle \in B$ three further conditions:

(forth) $\forall v_1' \in V_1 \forall a \in A \big( v_1 \xrightarrow{a}_1 v_1'$
$\qquad \implies \exists v_2' \in V_2 \big( v_2 \xrightarrow{a}_2 v_2' \land \langle v_1', v_2' \rangle \in B \big) \big)$,

(back) $\forall v_2' \in V_2 \forall a \in A$
$\qquad \big( \big( \exists v_1' \in V_1 \big( v_1 \xrightarrow{a}_1 v_1' \land \langle v_1', v_2' \rangle \in B \big) \big)$
$\qquad\qquad\qquad \Longleftarrow v_2 \xrightarrow{a}_2 v_2' \big)$,

(termination) $v_1\downarrow_1 \iff v_2\downarrow_2$.

For a function $f : V_1 \to V_2$ between $V_1$ and $V_2$ we say that $f$ *defines* a bisimulation between $\mathcal{C}_1$ and $\mathcal{C}_2$ if its graph $B_f := \{\langle v, f(v) \rangle \mid v \in V_1\}$ is a bisimulation between $\mathcal{C}_1$ and $\mathcal{C}_2$. Then we call $B_f$ a *functional* bisimulation. If $f$ is bijective, then $f$ is an *isomorphism* between $\mathcal{C}_1$ and $\mathcal{C}_2$.

We say that $\mathcal{C}_1$ and $\mathcal{C}_2$ are *bisimilar*, and denote it by $\mathcal{C}_1 \leftrightarrow \mathcal{C}_2$, if there is a bisimulation between $\mathcal{C}_1$ and $\mathcal{C}_2$. We denote by $\mathcal{C}_1 \rightrightarrows \mathcal{C}_2$ (and respectively by $\mathcal{C}_1 \simeq \mathcal{C}_2$) the stronger statement that there is a functional bisimulation between $\mathcal{C}_1$ and $\mathcal{C}_2$ (an isomorphism between $\mathcal{C}_1$ and $\mathcal{C}_2$).

**Definition 2.10.** Let $\mathcal{C} = \langle V, A, v_s, \rightarrow, \downarrow \rangle$ be a chart. By $\leftrightarrow_\mathcal{C}$ we denote *bisimilarity on* $\mathcal{C}$, the largest bisimulation (which is the union of all bisimulations) between $\mathcal{C}$ and $\mathcal{C}$ itself. If $w_1 \leftrightarrow_\mathcal{C} w_2$ holds for vertices $w_1, w_2 \in V$, then we say that $w_1$ *and* $w_2$ *are bisimilar in* $\mathcal{C}$. We call $\mathcal{C}$ a *bisimulation collapse* if $\leftrightarrow_\mathcal{C} = \mathrm{id}_V$ holds, that is, if bisimilar vertices of $\mathcal{C}$ are identical.

**Lemma 2.11.** *Every $1$-free star expression is also an under-star-$1$-free star expression, and hence $StExp^{(\mathbf{1})}(A) \subseteq StExp^{(*/\mathbf{1})}(A)$. The set $StExp^{(*/\mathbf{1})}(A)$ of under-star-$1$-free star expressions over $A$ is closed under taking derivatives.*

**Proposition 2.12.** *The image of the class $StExp^{(*/\mathbf{1})}$ of under-star-$1$-free star expressions via the chart interpretation $\mathcal{C}(\cdot)$ is closed under the operation of bisimulation collapse, modulo isomorphism.*

*Proof remark.* This statement can be shown with a refinement of the results and the methods for 1-free star expressions with binary star iteration in [10], [14] here for the class of under-star-1-free expressions that naturally extends the here corresponding class of 1-free star expressions with unary star iteration. For this result the careful formulation of the TSS in Def. 2.4 is crucial, see Rem. 2.6. □

**Example 2.13.** The chart below is the running example in [10]. It is a bisimulation collapse that satisfies LEE, as witnessed by the run of the loop elimination procedure in the middle, and the labeling that records the run on the right:

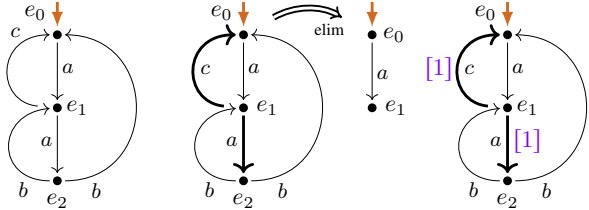

It is expressible by a star expression modulo bisimilarity [7]. Then by Prop. 2.12 it is isomorphic to the chart interpretation of a quasi-1-less star expression. Indeed, it is $\mathcal{C}(e_0)$ with start $e_0 = ((1 \cdot a) \cdot f) \cdot 0$ for $f = (c \cdot a + a \cdot (b + b \cdot a))^*$, and vertices $e_1 = (1 \cdot f) \cdot 0$, and $e_2 = ((1 \cdot (b + b \cdot a)) \cdot f) \cdot 0$.

As key finding in Thm. 5.8 we will show that the statement of Prop. 2.12 does not extend to all star expressions.

## 3. Refined process semantics

With the aim of making the structural property LEE useful for process interpretations of all star expressions (which can be violated as we have seen in Ex. 2.8), we introduce '1-charts' as charts with additional '1-transitions'. Based on that concept, we then define the '1-chart interpretation' that refines the chart interpretation, and produces 1-charts that satisfy LEE (as we will see in the next section).

**Definition 3.1.** A 1-*chart* is a 6-tuple $\langle V, A, 1, v_s, \rightarrow, \downarrow \rangle$ where $V$ is a set of *vertices*, $A$ is a set of *(proper) action labels*, $1 \notin A$ is the specified *empty step label*, $v_s \in V$ is the *start vertex* (hence $V \neq \varnothing$), $\rightarrow \subseteq V \times \underline{A} \times V$ is the *labeled transition relation*, where $\underline{A} := A \cup \{1\}$ is the set of action labels including 1, and $\downarrow \subseteq V$ is a set of *vertices with immediate termination*. (Note that then $\langle V, \underline{A}, v_s, \rightarrow, \downarrow \rangle$ is a chart). In such a 1-chart, we call a transition in $\rightarrow \cap (V \times A \times V)$ (labeled by a *proper action* in $A$) a *proper transition*, and a transition in $\rightarrow \cap (V \times \{1\} \times V)$ (labeled by the *empty-step symbol* 1) a 1-*transition*. Reserving non-underlined action labels like $a, b, \ldots$ for proper actions, we use underlined action label symbols like $\underline{a}, \underline{b}, \ldots$ for actions labels in the set $\underline{A}$ that includes the label 1. We highlight in red transition labels that may involve 1. We say that a 1-chart is 1-*free* if it does not contain 1-transitions.

Let $v \in V$ be a vertex in such a 1-chart $\underline{\mathcal{C}}$. We say that $v$ is a *proper-transition target (in $\underline{\mathcal{C}}$)* if it is the target of a proper transition of $\underline{\mathcal{C}}$. We say that $v$ is *start-vertex connected (in $\underline{\mathcal{C}}$)* if there is path of transitions from the start vertex $v_s$ of $\underline{\mathcal{C}}$ to $v$. We call $\underline{\mathcal{C}}$ *start-vertex connected* if every of its vertices is start-vertex connected.

**Definition 3.2.** Let $\underline{\mathcal{C}}_i = \langle V_i, A, 1, v_{s,i}, \rightarrow_i, \downarrow_i \rangle$ for $i \in \{1, 2\}$ be 1-charts over the same set $A$ of actions.

By a 1-*bisimulation between* $\underline{\mathcal{C}}_1$ *and* $\underline{\mathcal{C}}_2$ we mean a binary relation $B \subseteq V_1 \times V_2$ that is a bisimulation between the induced-transition charts $(\underline{\mathcal{C}}_1)_{(\cdot)}$ and $(\underline{\mathcal{C}}_2)_{(\cdot)}$.

We say that a partial function $f : V_1 \rightharpoonup V_2$ *defines* a 1-bisimulation between $\underline{\mathcal{C}}_1$ and $\underline{\mathcal{C}}_2$ if its graph $B_f := \{\langle v, f(v) \rangle \mid v \in V_1, f(v) \text{ defined}\}$, is a 1-bisimulation between $\underline{\mathcal{C}}_1$ and $\underline{\mathcal{C}}_2$. In this case we call $B_f$ a *functional* 1-bisimulation. If additionally $f$ is a bijective, total function, then we say that $f$ is an isomorphism between $\underline{\mathcal{C}}_1$ and $\underline{\mathcal{C}}_2$.

We write $\underline{\mathcal{C}}_1 \leftrightarrow \underline{\mathcal{C}}_2$, and say $\underline{\mathcal{C}}_1$ and $\underline{\mathcal{C}}_2$ *are* 1-*bisimilar*, if there is a 1-bisimulation between $\underline{\mathcal{C}}_1$ and $\underline{\mathcal{C}}_2$. We write $\underline{\mathcal{C}}_1 \rightrightarrows \underline{\mathcal{C}}_2$ if there is a functional 1-bisimulation between $\underline{\mathcal{C}}_1$ and $\underline{\mathcal{C}}_2$ ($\underline{\mathcal{C}}_1$ and $\underline{\mathcal{C}}_2$ *are functionally* 1-*bisimilar*). By $\underline{\mathcal{C}}_1 \simeq \underline{\mathcal{C}}_2$ we denote that there is an isomorphism between $\underline{\mathcal{C}}_1$ and $\underline{\mathcal{C}}_2$.

From every 1-chart a 1-bisimilar, start-vertex connected 1-chart can be obtained by garbage collection, that is, by removing all vertices that are not start-vertex connected.

**Convention 3.3.** By a (1-)chart we will henceforth mean a (1-)chart in which every vertex is start-vertex connected.

**Definition 3.4.** Let $\underline{\mathcal{C}} = \langle V, A, 1, v_s, \rightarrow, \downarrow \rangle$ be a 1-chart.

By $\leftrightarrow_{\underline{\mathcal{C}}}$ we denote 1-*bisimilarity on* $\underline{\mathcal{C}}$, the largest 1-bisimulation between $\underline{\mathcal{C}}$ and $\underline{\mathcal{C}}$ itself (which is the union of all 1-bisimulations between $\underline{\mathcal{C}}$ and $\underline{\mathcal{C}}$). If $w_1 \leftrightarrow_{\underline{\mathcal{C}}} w_2$ holds for vertices $w_1, w_2 \in V$, then we say that $w_1$ *and* $w_2$ *are* 1-*bisimilar in* $\underline{\mathcal{C}}$.

We call $\underline{\mathcal{C}}$ a 1-*bisimulation collapse* if $\leftrightarrow_{\underline{\mathcal{C}}} = \mathrm{id}_V$ holds, that is, if 1-bisimilar vertices of $\underline{\mathcal{C}}$ are identical. If, additionally, $\underline{\mathcal{C}}$ does not contain any 1-transitions, then we permit to call $\underline{\mathcal{C}}$ a *bisimulation collapse*.

Let $w_1, w_2 \in V$. We say that $w_1$ *is a substate of* $w_2$, denoted by $w_1 \sqsubseteq_{\underline{\mathcal{C}}} w_2$, if the pair $\langle w_1, w_2 \rangle$ forth-progresses to $1$-bisimilarity on $\underline{\mathcal{C}}$ in the sense of the following conditions:

(prog-forth) $\forall w_1' \in V_1 \forall a \in A \bigl( w_1 \xrightarrow{a} w_1'$
$\implies \exists w_2' \in V_2 \bigl( w_2 \xrightarrow{a} w_2' \wedge w_1' \Leftrightarrow_{\underline{\mathcal{C}}} w_2' \bigr) \bigr),$
(prog-termination) $w_1 \downarrow \implies w_2 \downarrow$.

**Definition 3.5.** We say that a $1$-chart $\underline{\mathcal{C}}$ is *weakly guarded* (w.g.) if $\underline{\mathcal{C}}$ does not contain an infinite path of $1$-transitions.

**Definition 3.6.** The $1$-*chart interpretation of* a star expression $e \in \mathit{StExp}(A)$ is the $1$-chart of the form:

$$\underline{\mathcal{C}}(e) = \langle \underline{V}(e), A, 1, e, \rightarrow \cap (\underline{V}(e) \times \underline{A} \times \underline{V}(e)), \downarrow \cap \underline{V}(e) \rangle,$$

which is based on $\rightarrow \, \subseteq \mathit{StExp}(A) \times \underline{A} \times \mathit{StExp}(A)$, the the labeled transition relation that is defined via derivability in the transition system specification $\underline{\mathcal{T}}(A)$ (where $a \in A$, $\underline{a} \in \underline{A} = A \cup \{1\}$, and $e, e_1, e_2 \in \mathit{StExp}(A)$):

$$\overline{a \xrightarrow{a} 1} \qquad \overline{e_1 + e_2 \xrightarrow{1} e_i}$$

$$\frac{e_1 \xrightarrow{\underline{a}} e_1'}{e_1 \cdot e_2 \xrightarrow{\underline{a}} e_1' \cdot e_2} \text{ (if } e_1' \text{ is normed)} \qquad \frac{e_1 \xrightarrow{\underline{a}} e_1'}{e_1 \cdot e_2 \xrightarrow{\underline{a}} e_1'} \text{ (if } e_1' \text{ is not normed)}$$

$$\overline{1 \cdot e_2 \xrightarrow{1} e_2} \qquad \overline{e^* \xrightarrow{1} 1}$$

$$\frac{e \xrightarrow{1}{}^* \cdot \xrightarrow{a} e'}{e^* \xrightarrow{a} e' \cdot e^*} \text{ (if } e' \text{ is normed)} \qquad \frac{e \xrightarrow{1}{}^* \cdot \xrightarrow{a} e'}{e^* \xrightarrow{a} e'} \text{ (if } e' \text{ is not normed)}$$

such that $\underline{V}(e)$ is defined as the set of those star expressions that are reachable from $e$ via transitions of $\rightarrow$, from which the set $\rightarrow \cap (\underline{V}(e) \times \underline{A} \times \underline{V}(e))$ of transitions of $\underline{\mathcal{C}}(e)$ is defined, and the set $\downarrow \cap \underline{V}(e)$ of terminating vertices by using $\downarrow := \{1\}$ (so here only $1$ is be a terminating vertex if it is reachable).

**Example 3.7.** For the star expression $e = (a^* \cdot b^*)^*$, the $1$-chart interpretation $\underline{\mathcal{C}}(e)$ as obtained by the TSS $\underline{\mathcal{T}}(A)$ is:

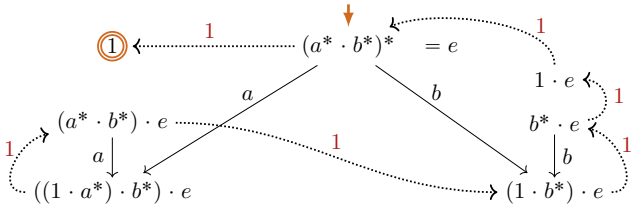

Note that $1$ is the single vertex with immediate termination, and that the $1$-chart $\underline{\mathcal{C}}(e)$ is weakly guarded. Neglecting the actions, the form of $\underline{\mathcal{C}}(e)$ corresponds to that of the graph $G_3$ on page 1. This implies, by using an analogous run of the loop subchart elimination procedure for $\underline{\mathcal{C}}(e)$ as the one we illustrated for $G_3$ on page 2, that also $\underline{\mathcal{C}}(e)$ satisfies LEE.

**Lemma 3.8.** *Derivability of statements concerning termination, and transitions in $\mathcal{T}(A)$, and in $\underline{\mathcal{T}}(A)$ are related as follows, for all $e, e' \in \mathit{StExp}(A)$, and $a \in A$:*

$$\vdash_{\mathcal{T}} e \downarrow \iff \vdash_{\underline{\mathcal{T}}} e \xrightarrow{1}{}^* 1, \tag{3.1}$$

$$\vdash_{\mathcal{T}} e \xrightarrow{a} e' \iff \vdash_{\underline{\mathcal{T}}} e \xrightarrow{1}{}^* \cdot \xrightarrow{a} e'. \tag{3.2}$$

**Lemma 3.9.** *For every $e \in \mathit{StExp}(A)$, the $1$-chart interpretation $\underline{\mathcal{C}}(e)$ of $e$ is a finite, weakly guarded $1$-chart.*

**Definition 3.10.** Let $\underline{\mathcal{C}} = \langle V, A, 1, v_s, \rightarrow, \downarrow \rangle$ be a $1$-chart. By the *induced chart of* $\underline{\mathcal{C}}$, and the *chart induced by* $\underline{\mathcal{C}}$, we mean the $1$-chart $\underline{\mathcal{C}}_{(\cdot)} = \langle V_0, A, 1, v_s, \xrightarrow{(\cdot)} \cap (V_0 \times A \times V_0), \downarrow^{(1)} \cap V_0 \rangle$ where $\xrightarrow{(\cdot)} \, \subseteq V \times A \times V$ is the *induced transition relation*, and $\downarrow^{(1)} \subseteq V$ is the set of vertices with *induced termination* that are defined as follows, for all $v, v' \in V$ and $a \in A$:

(ind-1) $v \xrightarrow{(a)} v'$ holds if $v = v_0 \xrightarrow{1} v_1 \xrightarrow{1} \ldots \xrightarrow{1} v_n \xrightarrow{a} v'$, for some $v_0, \ldots, v_n \in V$ and $n \in \mathbb{N}$ (we say there is an *induced transition* between $v, v' \in V$ w.r.t. $\underline{\mathcal{L}}$),

(ind-2) $v \downarrow^{(1)}$ holds if $v = v_0 \xrightarrow{1} v_1 \xrightarrow{1} \ldots \xrightarrow{1} v_n \wedge v_n \downarrow^{(1)}$, for some $v_0, \ldots, v_n \in V$ and $n \in \mathbb{N}$ (then we say that $v$ *has induced termination with respect to* $\underline{\mathcal{L}}$).

and where $V_0 \subseteq V$ is the set of all vertices that are reachable from $v_s$ by induced transitions. The notation $\xrightarrow{(a)}$ intends to reflect the asymmetry that an induced $a$-transition consists of an arbitrary number of leading $1$-transitions that is trailed by a single proper $a$-transition.

**Theorem 3.11.** $\underline{\mathcal{C}}(e)_{(\cdot)} = \mathcal{C}(e)$ *for all $e \in \mathit{StExp}(A)$, that is, the chart interpretation $\mathcal{C}(e)$ of a star expression $e$ is the induced chart of the $1$-chart interpretation $\underline{\mathcal{C}}(e)$ of $e$.*

*Proof.* Lem. 3.8 shows, for $e \in \mathit{StExp}(A)$, that termination in $\mathcal{C}(e)$ coincides with induced termination in $\underline{\mathcal{C}}(e)$ (due to (3.1) since $\underline{\mathcal{C}}(e)$ can only have $1$ as terminating vertex), and that transitions in $\mathcal{C}(e)$ coincide with induced transitions in $\underline{\mathcal{C}}(e)$ (by (3.2)). This entails $\underline{\mathcal{C}}(e)_{(\cdot)} = \mathcal{C}(e)$. □

**Example 3.12.** The picture below can explain that the chart interpretation $\mathcal{C}(e)$ of $e = (a^* \cdot b^*)^*$ in Ex. 2.7 is the induced chart of the $1$-chart interpretation $\underline{\mathcal{C}}(e)$ in Ex. 3.7:

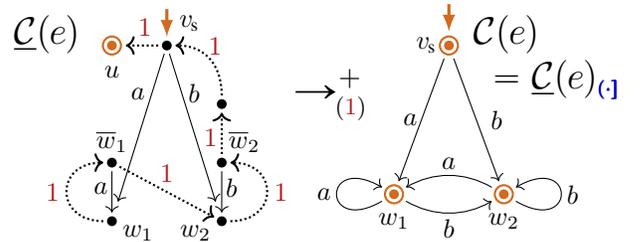

Note that, for instance, $w_1$ has induced termination in $\underline{\mathcal{C}}(e)$, because there is a $1$-transition path from $w_1$ via $\overline{w}_1, w_2, \overline{w}_2$, and $v_s$ to the terminating vertex $u$, which gives rise to $w_1$ being terminating in $\underline{\mathcal{C}}(e)_{(\cdot)}$. Similarly, the $b$-transition from $w_1$ to $w_2$ in $\underline{\mathcal{C}}(e)_{(\cdot)}$ arises from an induced $b$-transition in $\underline{\mathcal{C}}(e)$ that consist of the $1$-transition path from $w_1$ to $v_s$ which is then followed by the proper $b$-transition from $v_s$ to $w_2$.

We have added the symbol $\rightarrow^+_{(1)}$ for a rewrite sequence of the relation $\rightarrow_{(1)}$ in anticipation of its definition below.

**Definition 3.13.** Let $\underline{\mathcal{C}}_i = \langle V_i, A, 1, v_s, \rightarrow_i, \downarrow_i \rangle$ for $i \in \{1, 2\}$ be $1$-charts. We denote by $\underline{\mathcal{C}}_1 \rightarrow_{(1)} \underline{\mathcal{C}}_2$ that $\underline{\mathcal{C}}_2$ arises from $\underline{\mathcal{C}}_1$ by a $1$-*transition elimination step* according to one of two local rules each of which removes the $1$-transition $\langle v_0, 1, v \rangle$:

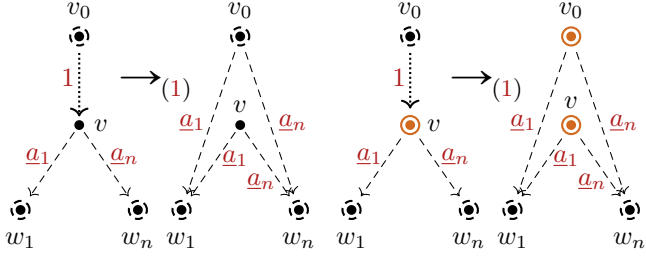

and adds, for each transition $\langle v, \underline{a_i}, w_i \rangle$ with $i \in \{1, \ldots, n\}$ from $v$, a cofinal transition $\langle v_0, \underline{a_i}, w_i \rangle$ with the same label from $v$. Note that $n = 0$ is possible if $v$ has outdegree 0. Additionally, garbage collection of $v$ is permitted if $v$ becomes unreachable. That immediate termination is permitted (but not required) in a vertex is indicated by a dashed outer ring. While $v$ does not permit immediate termination in the rule on the left, immediate termination at $v$ in the rule on the right is transferred in the step to $v_0$. We used dashed arrows for the outgoing transitions from $v$ to indicate that they can be either proper transitions or 1-transitions.

We say that $\underline{\mathcal{C}}_1$ (1-transition) refines $\underline{\mathcal{C}}_2$, that $\underline{\mathcal{C}}_1$ is a (1-transition) refinement of $\underline{\mathcal{C}}_2$, and that $\underline{\mathcal{C}}_2$ is or can be (1-transition) refined by $\underline{\mathcal{C}}_1$, if $\underline{\mathcal{C}}_1 \to^*_{(1)} \underline{\mathcal{C}}_2$, that is, $\underline{\mathcal{C}}_2$ arises from $\underline{\mathcal{C}}_1$ by a sequence of 1-transition elimination steps.

**Remark 3.14.** We will only be concerned with the elimination of 1-transitions from finite, weakly guarded 1-charts. Therefore we could have restricted the two rules in Def. 3.13 by demanding (i) $v_0 \neq v$ (since w.g. 1-charts do not have 1-transition self-loops), and (ii) that all outgoing transitions from $v$ are proper transitions (because all 1-transitions in finite w.g. 1-charts can be eliminated in a bottomup manner). We left out restriction (i) for simplicity, thereby accepting that it introduces a cyclic reduction for 1-transition selfloops for not w.g. 1-charts. But we needed to avoid restriction (ii), because in Section 4 we will use annotated versions of these rules for situations when it is not possible to eliminate all 1-transitions while keeping the property LEE.

**Lemma 3.15.** *The 1-transition elimination rewrite relation $\to_{(1)}$ has the following properties, for all 1-charts $\underline{\mathcal{C}}, \underline{\mathcal{C}}_1, \underline{\mathcal{C}}_2$:*

*(i) If $\underline{\mathcal{C}}_1$ is weakly guarded, and $\underline{\mathcal{C}}_1 \to^*_{(1)} \underline{\mathcal{C}}_2$, then $\mathcal{C}_2$ is finite and w.g., and $|V(\mathcal{C}_1)| - 1 \leq |V(\mathcal{C}_2)| \leq |V(\mathcal{C}_1)|$.*

*(ii) $\to_{(1)}$ is terminating from every finite w.g. 1-chart.*

*(iii) $\to_{(1)}$ normal forms are 1-free 1-chart. $\to_{(1)}$ normal forms of finite, w.g. 1-charts are finite 1-free 1-charts.*

*(iv) $\underline{\mathcal{C}}_1 \to^*_{(1)} \underline{\mathcal{C}}_2 \implies (\underline{\mathcal{C}}_1)_{(\cdot]} = (\underline{\mathcal{C}}_2)_{(\cdot]}$.*

*(v) If $\underline{\mathcal{C}}$ is finite and weakly guarded, and $\mathcal{C}$ is 1-free, then: $\underline{\mathcal{C}} \to^*_{(1)} \mathcal{C} \iff (\underline{\mathcal{C}})_{(\cdot]} = \mathcal{C}$.*

*(vi) If $\underline{\mathcal{C}}$ is finite, and weakly guarded, then $\underline{\mathcal{C}} \to^*_{(1)} \underline{\mathcal{C}}_{(\cdot]}$, that is, $\underline{\mathcal{C}}$ refines its induced chart $\underline{\mathcal{C}}_{(\cdot]}$, and $\underline{\mathcal{C}}_{(\cdot]}$ is the unique $\to_{(1)}$ normal form of $\underline{\mathcal{C}}$.*

**Corollary 3.16.** $\underline{\mathcal{C}}(e) \to^*_{(1)} \underline{\mathcal{C}}(e)_{(\cdot]} = \mathcal{C}(e)$ *for every star expression $e \in StExp(A)$, i.e. the 1-chart interpretation $\underline{\mathcal{C}}(e)$ of $e$ refines the chart interpretation $\mathcal{C}(e)$ of $e$.*

*Proof.* For $e \in StExp(A)$, $\underline{\mathcal{C}}(e)_{(\cdot]} = \mathcal{C}(e)$ by Thm. 3.11, and $\underline{\mathcal{C}}(e)$ is finite and weakly guarded by Lem. 3.9. Then by Lem. 3.15, (vi), we obtain $\underline{\mathcal{C}}(e) \to^*_{(1)} \underline{\mathcal{C}}(e)_{(\cdot]} = \mathcal{C}(e)$. $\square$

## 4. LLEE-witnesses for the refined semantics

In Ex. 3.7 we noticed that the 1-chart interpretation $\underline{\mathcal{C}}(e)$ of $e = (a^* \cdot b^*)^*$ permits a similar run of the loop elimination procedure as that for $G_3$ on page 2, and that therefore also $\underline{\mathcal{C}}(e)$ satisfies LEE. This successful run of the loop elimination procedure can be recorded on $\underline{\mathcal{C}}(e)$ by attaching to a transition $\tau$ of $\underline{\mathcal{C}}(e)$ the marking label $[n]$ for $n \in \{1, 2, 3\}$ if $\tau$ is eliminated in the $n$-th step, thereby obtaining the labeled version $L_1$ below (where we neglect the action labels):

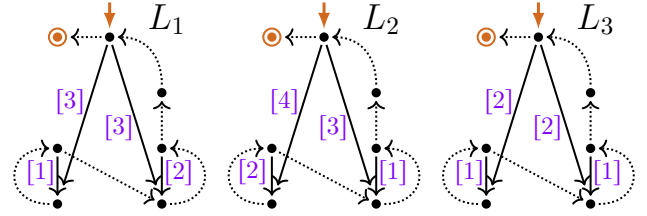

The labelings $L_2$ and $L_3$ of $\underline{\mathcal{C}}(e)$ record two other successful runs of the loop elimination procedure of length 4 and 2, respectively, where for $L_3$ we have permitted to eliminate two loop subcharts at different vertices together in the first step. Such labelings of successful runs of the loop elimination procedure we call 'LEE-witnesses', and indeed 'layered LLEE-witnesses', because in the runs recorded in $L_1$, $L_2$, and $L_3$ loop-entry transitions are never removed from the body of a previously removed loop subcharts. This restriction of the elimination procedure, which can be shown to not affect the property LEE, leads to a concept that is easier to reason about, and that still suffices for our purposes.

Following the development in [10], we define 'layered LEE-witnesses' of a 1-chart $\underline{\mathcal{C}}$ as 'entry/body-labelings' of $\underline{\mathcal{C}}$ that delimit a hierarchical structure of 'loop 1-charts' in $\underline{\mathcal{C}}$. But we change the condition (L3) for 'loop charts' in [10], which excluded terminating vertices, to permit immediate termination at the start vertex of 'loop 1-charts'. In dealing with star expressions like $1 \cdot a^*$ that are not 1-free, we want to call the chart interpretation $\mathcal{C}(1 \cdot a^*)$ (which consists of a terminating start vertex with an $a$-self-loop) a 'loop 1-chart'.

**Definition 4.1.** A 1-chart $\underline{\mathcal{L}} = \langle V, A, 1, v_s, \to, \downarrow \rangle$ is called a *loop 1-chart* if it satisfies the following three conditions:

(L1) There is an infinite path from the start vertex $v_s$.

(L2) Every infinite path from $v_s$ returns to $v_s$ after a positive number of transitions (and so visits $v_s$ infinitely often).

(L3) Immediate termination is permitted (but not required) only at the start vertex, that is, $\downarrow \subseteq \{v_s\}$.

We call the transitions from $v_s$ *loop-entry transitions*, and all other transitions *loop-body transitions*. A loop 1-chart $\underline{\mathcal{L}}$ is a *loop sub-1-chart of* a 1-chart $\underline{\mathcal{C}}$ if it is the sub-1-chart of $\underline{\mathcal{C}}$ rooted at some vertex $v \in V$ that is generated, for a nonempty set $U$ of transitions of $\mathcal{C}$ from $v$, by all paths that start with a transition in $U$, and continue until $v$ is reached again (then $U$ is the set of loop-entry transitions of $\underline{\mathcal{L}}$).

**Definition 4.2.** By an *entry/body-labeling of a* 1-*chart* $\underline{\mathcal{C}} = \langle V, A, 1, v_s, \rightarrow, \downarrow \rangle$ we mean a chart $\hat{\underline{\mathcal{C}}} = \langle V, \underline{A} \times \mathbb{N}, v_s, \hat{\rightarrow}, \downarrow \rangle$ where $\hat{\rightarrow} \subseteq V \times (\underline{A} \times \mathbb{N}) \times V$ arises from $\rightarrow$ by adding, for each transition $\tau = \langle v_1, \underline{a}, v_2 \rangle \in \rightarrow$, to the action label $\underline{a}$ of $\tau$ a *marking label* $n \in \mathbb{N}$, yielding $\hat{\tau} = \langle v_1, \langle \underline{a}, n \rangle, v_2 \rangle \in \hat{\rightarrow}$. In an entry/body-labeling we call transitions with marking label 0 *body transitions*, and transitions with marking labels in $\mathbb{N}^+$ *entry transitions*.

Let $\hat{\underline{\mathcal{C}}}$ be an entry/body-labeling of $\underline{\mathcal{C}}$, and let $v$ and $w$ be vertices of $\underline{\mathcal{C}}$ and $\hat{\underline{\mathcal{C}}}$. We denote by $v \rightarrow_{\mathsf{bo}} w$ that there is a body transition $v \xrightarrow{\langle \underline{a}, 0 \rangle} w$ in $\hat{\underline{\mathcal{C}}}$ for some $\underline{a} \in \underline{A}$, and by $v \rightarrow_{[n]} w$, for $n \in \mathbb{N}^+$ that there is an entry transition $v \xrightarrow{\langle \underline{a}, n \rangle} w$ in $\hat{\underline{\mathcal{C}}}$ for some $\underline{a} \in \underline{A}$. By $E(\hat{\underline{\mathcal{C}}})$ of *entry transition identifiers* we denote the set of pairs $\langle v, n \rangle \in V \times \mathbb{N}^+$ such that an entry transition $\rightarrow_{[n]}$ departs from $v$ in $\hat{\underline{\mathcal{C}}}$. For $\langle v, n \rangle \in E(\hat{\underline{\mathcal{C}}})$, we define by $\underline{\mathcal{C}}_{\hat{\underline{\mathcal{C}}}}(v, n)$ the subchart of $\underline{\mathcal{C}}$ with start vertex $v_s$ that consists of the vertices and transitions which occur on paths in $\underline{\mathcal{C}}$ as follows: any path that starts with a $\rightarrow_{[n]}$ entry transition from $v$, continues with body transitions only (thus does not cross another entry transition), and halts immediately if $v$ is revisited. By a *back-link* of $\hat{\underline{\mathcal{C}}}$ we mean a transition of $\hat{\underline{\mathcal{C}}}$ back to the start vertex $v$ of an induced sub-1-chart $\underline{\mathcal{C}}_{\hat{\underline{\mathcal{C}}}}(v, n)$ of $\hat{\underline{\mathcal{C}}}$, for some $\langle v, n \rangle \in E(\hat{\underline{\mathcal{C}}})$.

**Definition 4.3.** Let $\underline{\mathcal{C}} = \langle V, A, v_s, 1, \rightarrow, \downarrow \rangle$ be a 1-chart. A LLEE-*witness* (a *layered* LEE-*witness*) *of* $\underline{\mathcal{C}}$ is an entry/body-labeling $\hat{\underline{\mathcal{C}}}$ of $\underline{\mathcal{C}}$ with the following three properties:

(W1) *Body-step termination:* There is no infinite path of $\rightarrow_{\mathsf{bo}}$ transitions in $\underline{\mathcal{C}}$.
(W2) *Loop condition:* For all $\langle v, n \rangle \in E(\hat{\underline{\mathcal{C}}})$, the 1-chart $\underline{\mathcal{C}}_{\hat{\underline{\mathcal{C}}}}(v, n)$ is a loop 1-chart.
(W3) *Layeredness:* For all $\langle v, n \rangle \in E(\hat{\underline{\mathcal{C}}})$, if an entry transition $w \rightarrow_{[m]} w'$ departs from a state $w \neq v$ of $\underline{\mathcal{C}}_{\hat{\underline{\mathcal{C}}}}(v, n)$, then its marking label $m$ satisfies $m < n$.

The condition (W2) justifies to call an entry transition in a LLEE-witness a *loop-entry transition*. For a loop-entry transition $\rightarrow_{[m]}$ with $m \in \mathbb{N}^+$, we call $m$ its *loop level*.

We call $\underline{\mathcal{C}}$ a LLEE-1-*chart* if it has a LLEE-witness, and a wg-LLEE-1-*chart* if, additionally, it is weakly guarded.

**Proposition 4.4.** *Every* LLEE-1-*chart satisfies* LEE.

*Proof.* Let $\hat{\underline{\mathcal{C}}}$ be a LLEE-witness of a 1-chart $\underline{\mathcal{C}}$. Repeatedly pick an entry transition identifier $\langle v, n \rangle \in E(\hat{\underline{\mathcal{C}}})$ with $n \in \mathbb{N}^+$ minimal, remove the loop sub-1-chart $\underline{\mathcal{C}}_v(n)$ that is generated by loop-entry transitions of level $n$ from $v$ (it is indeed a loop by condition (W2) on $\hat{\underline{\mathcal{C}}}$, noting that minimality of $n$ and condition (W3) on $\hat{\underline{\mathcal{C}}}$ ensure the absence of departing loop-entry transitions of lower level), and perform garbage collection. Eventually the part of $\mathcal{C}$ that is reachable by body transitions from the start vertex is obtained. This sub-1-chart of $\underline{\mathcal{C}}$ does not have an infinite path due to condition (W1) on $\hat{\underline{\mathcal{C}}}$. Therefore $\underline{\mathcal{C}}$ satisfies LEE. □

**Definition 4.5.** We denote by $\hat{\mathcal{T}}(A)$ the TSS that is a marking labeled version $\mathcal{T}(A)$ in Def. 3.6 (where $a \in A$, $\underline{a} \in \underline{A}$, $e, e_1, e_2 \in StExp(A)$, and $l \in \mathbb{N}$ are arbitrary):

$$\frac{}{a \xrightarrow{\underline{a}}_{\mathsf{bo}} 1} \qquad \frac{}{e_1 + e_2 \xrightarrow{1}_{\mathsf{bo}} e_i}$$

$$\frac{e_1 \xrightarrow{\underline{a}}_l e_1'}{e_1 \cdot e_2 \xrightarrow{\underline{a}}_l e_1' \cdot e_2} \text{ (if } e_1' \text{ is normed)} \qquad \frac{e_1 \xrightarrow{\underline{a}}_l e_1'}{e_1 \cdot e_2 \xrightarrow{\underline{a}}_{\mathsf{bo}} e_1'} \text{ (if } e_1' \text{ is not normed)}$$

$$\frac{}{1 \cdot e_2 \xrightarrow{1}_{\mathsf{bo}} e_2} \qquad \frac{}{e^* \xrightarrow{1}_{\mathsf{bo}} 1}$$

$$\frac{e \xrightarrow{1}_{\mathsf{bo}}^* \cdot \xrightarrow{\underline{a}}_l e'}{e^* \xrightarrow{\underline{a}}_{[|e^*|_*]} e' \cdot e^*} \text{ (if } e' \text{ is normed)} \qquad \frac{e \xrightarrow{1}_{\mathsf{bo}}^* \cdot \xrightarrow{\underline{a}}_l e'}{e^* \xrightarrow{\underline{a}}_{\mathsf{bo}} e'} \text{ (if } e' \text{ is not normed)}$$

Derivability in $\hat{\mathcal{T}}$ defines the marking labeled version $\hat{\rightarrow} \subseteq StExp(A) \times (\underline{A} \times \mathbb{N}) \times StExp(A)$ of the transition relation $\rightarrow \subseteq StExp(A) \times \underline{A} \times StExp(A)$ as defined by $\mathcal{T}(A)$.

The *entry/body-labeling* $\widehat{\underline{\mathcal{C}}(e)}$ of the 1-chart interpretation $\underline{\mathcal{C}}(e)$ of a star expression $e \in StExp(A)$ is defined as:

$$\widehat{\underline{\mathcal{C}}(e)} = \langle \underline{V}(e), \underline{A}, e, \hat{\rightarrow} \cap (\underline{V}(e) \times (\underline{A} \times \mathbb{N}) \times \underline{V}(e)), \downarrow \cap \underline{V}(e) \rangle,$$

where $\underline{V}(e)$ is the set of star expressions that are reachable from $e$ via transitions of $\hat{\rightarrow}$ (or $\rightarrow$), from which the set $\rightarrow \cap (\underline{V}(e) \times \underline{A} \times \underline{V}(e))$ of transitions of $\underline{\mathcal{C}}(e)$ was defined, and the set $\downarrow \cap \underline{V}(e)$ of terminating vertices of $\underline{\mathcal{C}}(e)$ by $\downarrow := \{1\}$.

**Example 4.6.** For $e = (a^* \cdot b^*)^*$ we obtain the entry/body-labeling $\widehat{\underline{\mathcal{C}}(e)}$ as defined by the TSS $\hat{\mathcal{T}}(A)$ can be obtained similarly as we obtained the 1-chart interpretation $\underline{\mathcal{C}}(e)$ of $e$ in Ex. 3.7. By dropping the action labels we obtain:

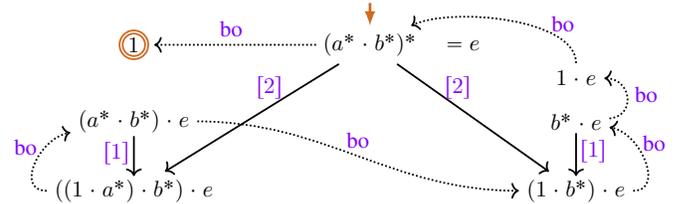

Here and below we draw loop-entry transitions as thicker arrows. $\widehat{\underline{\mathcal{C}}(e)}$ obviously is an entry/body-labeling of $\underline{\mathcal{C}}(e)$, and it is also easy to verify that $\widehat{\underline{\mathcal{C}}(e)}$ is a LLEE-witness of $e$.

Complementing Thm. 3.11, we now formulate the result that the 1-chart interpretation of a star expression refines the chart interpretation in such a way that LEE is recovered.

**Theorem 4.7.** *The entry/body-labeling* $\widehat{\underline{\mathcal{C}}(e)}$ *of* $\underline{\mathcal{C}}(e)$ *is a* LLEE-*witness of the* 1-*chart interpretation* $\underline{\mathcal{C}}(e)$ *of* $e$, *for every* $e \in StExp(A)$. *Therefore the* 1-*chart interpretation* $\underline{\mathcal{C}}(e)$ *of a star expression* $e \in StExp(A)$ *satisfies* LEE.

**Definition 4.8.** Let $\underline{\mathcal{C}}$ be a 1-chart. We say that $\underline{\mathcal{C}}$ *is* 1-*transition limited with respect to* a LLEE-witness $\hat{\underline{\mathcal{C}}}$ of $\underline{\mathcal{C}}$ if every 1-transition of $\underline{\mathcal{C}}$ lifts to a *back-link* in $\hat{\underline{\mathcal{C}}}$. We say that $\underline{\mathcal{C}}$ *is* 1-*transition limited* if $\underline{\mathcal{C}}$ is 1-transition limited with respect to some LLEE-witness $\hat{\underline{\mathcal{C}}}$ of $\underline{\mathcal{C}}$.

We say that $\underline{\mathcal{C}}$ *has the property* LLEE–1-lim if $\underline{\mathcal{C}}$ is weakly guarded, satisfies LEE, and is 1-transition limited.

**Lemma 4.9.** *Every finite wg-LLEE-1-chart reduces via a sequence of $\to_{(1)}$ steps to a finite 1-chart with LLEE–1-lim.*

For the proof we annotate the two 1-transition elimination rules in Def. 3.13 with marking labels $l, l_1, \ldots, l_n \in \mathbb{N}$ if the 1-transition from $v_0$ to $v$ is not a back-link (denoted by $\neg(v_0 \curvearrowright v)$) as follows (we neglect the action labels):

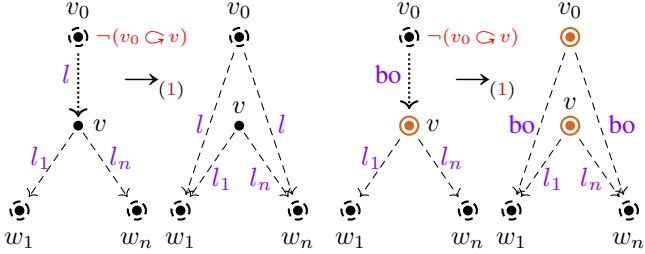

*Proof of Lem. 4.9.* It suffices to show that in a LLEE-witness $\hat{\underline{\mathcal{C}}}$ of a weakly guarded 1-chart $\underline{\mathcal{C}}$ every 1-transition that is not a back-link can be eliminated. Applications of the two annotated 1-transition elimination rules as above transform LLEE-witnesses and underlying 1-charts, and their repeated use leads to the elimination of all 1-transitions that are not back-links. □

**Example 4.10.** The LLEE-witness $\widehat{\underline{\mathcal{C}}(e)}$ for $e = (a^* \cdot b^*)^*$ in Ex. 4.6 is not 1-transition limited. It can be transformed into a 1-transition limited LLEE-witness by four steps according to the annotated $\to_{(1)}$ rules in the proof of Lem. 4.9:

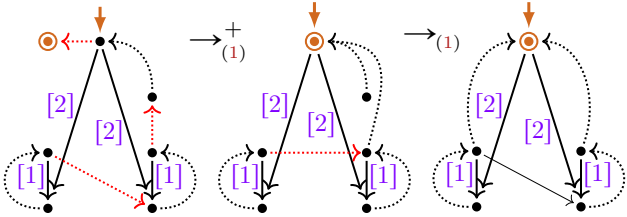

where in the first transformation above three $\to_{(1)}$ steps are performed in parallel to the 1-transitions highlighted in red. Here and below we drop the marking labels of body transitions, but instead contrast them by emphasizing the loop-entry transitions as thick arrows together with their levels.

**Remark 4.11.** Another variation $\underline{\mathcal{C}}'(\cdot)$ of the chart interpretation $\mathcal{C}(e)$ that interprets star expressions $e$ as 1-charts $\underline{\mathcal{C}}'(e)$ which are 1-bisimilar to $\mathcal{C}(e)$ and $\underline{\mathcal{C}}(e)$ is defined in [13]. That variation $\underline{\mathcal{C}}'(\cdot)$ produces finite, weakly guarded LLEE-1-charts and associated LLEE-witnesses that are 1-transition limited directly, without the need to apply Lem. 4.9 afterwards. But interpretations $\underline{\mathcal{C}}'(e)$ are only functionally 1-bisimilar to chart interpretations $\mathcal{C}(e)$, and they are *not* refinements of $\mathcal{C}(e)$. Also, the definition of $\underline{\mathcal{C}}'(\cdot)$ is based on the extension of star expressions by an additional operation 'stack product' $\circledast$ that keeps track of whether a derivative has originated by steps into an iteration. That device is used to ensure that 1-transitions are only introduced as back-links of derivatives to iterations from which they originated.

**Lemma 4.12.** *Let $\mathcal{C}$ be a finite 1-free 1-chart. If $\mathcal{C}$ can be refined into a wg-LLEE-1-chart, then $\mathcal{C}$ can be refined into a 1-chart with LLEE–1-lim.*

*Proof.* Suppose that $\underline{\mathcal{C}} \to^*_{(1)} \mathcal{C}$ for some wg-LLEE-1-chart $\underline{\mathcal{C}}$. Then $\underline{\mathcal{C}}$ is finite, due to Lem. 3.15, (i). Then $\underline{\mathcal{C}}_{(\cdot]} = \mathcal{C}$ by Lem. 3.15, (ii), since $\mathcal{C}$ is 1-free. Now we can also apply Lem. 4.9 to $\underline{\mathcal{C}}$ to find a 1-chart $\underline{\mathcal{C}}'$ with LLEE–1-lim such that $\underline{\mathcal{C}} \to^*_{(1)} \underline{\mathcal{C}}'$. Then we get $\underline{\mathcal{C}}'_{(\cdot]} = \underline{\mathcal{C}}_{(\cdot]} = \mathcal{C}$ by Lem. 3.15, (iv), and subsequently $\underline{\mathcal{C}}' \to^*_{(1)} \mathcal{C}$ by Lem. 3.15, (v). In this way we have shown that $\mathcal{C}$ can be refined into the 1-chart $\underline{\mathcal{C}}'$ with LLEE–1-lim. □

## 5. Counterexample

In this section we develop, in steps, the counterexample for the statements **(RC)** and **(IC)$_1$** on page 3 in Sect. 1. We start by immediately presenting the not collapsed LLEE-1-chart on which all of our observations will be based.

**Example 5.1.** We consider the 1-chart $\underline{\mathcal{C}}$ with actions in $\{a, a_1, a_2, b, c, c_1, c_2, d, e, f\}$, and the LLEE-witness $\hat{\underline{\mathcal{C}}}$ of $\underline{\mathcal{C}}$ that is indicated by colored loop-entry transitions from vertex $v$ (of level 2, blue) and from vertex $\overline{w}_1$ (of level 1, green):

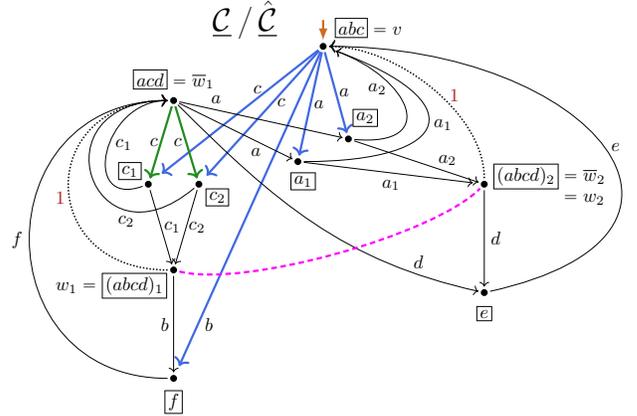

The framebox name of a vertex $u$ symbolizes the behavior from $u$ in $\underline{\mathcal{C}}$ by listing the actions of induced transitions from $u$. This is meaningful because all transitions with the same label from $\{b, d\}$ are cofinal, respectively, transitions with the same label in $\{a_1, a_2, c_1, c_2, e, f\}$ depart from just a single vertex, respectively, and transitions with the same label in $\{a, c\}$ occur as pairs of co-initial transitions such that the transitions of such a pair from a vertex $u_1$ can be joined in their targets by the transitions of the corresponding pair from a vertex $u_2$. It follows that vertices with the same label list are 1-bisimilar. In particular, the vertex pair $w_1 = \boxed{(abcd)_1}$ and $w_2 = \boxed{(abcd)_2}$ (their action lists are indexed to distinguish them) are 1-bisimilar (indicated by the line in magenta): every induced transition from $w_1$ can be joined by a cofinal induced transition from $w_2$, and vice versa. Thus $\underline{\mathcal{C}}$ is not a 1-bisimulation collapse.

**Lemma 5.2.** *The 1-chart $\underline{\mathcal{C}}$ in Ex. 5.1 is a LLEE-1-chart. Its induced chart $\underline{\mathcal{C}}_{(\cdot]}$ is isomorphic to the chart interpretation of a star expression.*

*Proof.* It is easy to check that the entry/body-labeling $\hat{\mathcal{C}}$ indicated in Ex. 5.1 is a LLEE-witness of $\underline{\mathcal{C}}$. It is also routine to verify that $\underline{\mathcal{C}}_{(\cdot]} \simeq \mathcal{C}(g_v)$ for the star expression $g_v \in \mathit{StExp}(\{a, a_1, a_2, b, c, c_1, c_2, d, e, f\})$ that is defined as:

$$g_v := (1 \cdot ((C + b \cdot f) \cdot g_{\overline{w}_1} + A)^*) \cdot 0$$
$$\text{where: } C := c \cdot (c_1 + c_1 \cdot g_{w_1}) + c \cdot (c_2 + c_2 \cdot g_{w_1})$$
$$g_{w_1} := 1 + b \cdot f$$
$$g_{\overline{w}_1} := (1 \cdot C^*) \cdot (A + d \cdot e)$$
$$A := a \cdot (a_1 + a_1 \cdot g_{w_2}) + a \cdot (a_2 + a_2 \cdot g_{w_2})$$
$$g_{w_2} := 1 + d \cdot e$$

Hereby the meaning of a star expression $g_u$ for some vertex $u$ of $\underline{\mathcal{C}}$ is to express the behavior in $\underline{\mathcal{C}}$ from vertex $u$ until for the first time a vertex is reached to which $u$ directly loops back (which for $u = v$ coincides with the behavior from $u$ as $v$ does not loop back). $A$ and $C$ express the behavior in $\underline{\mathcal{C}}$ starting from a vertex with the pair of $a$- and $c$-transitions, respectively, until the vertex is reached to which both of their targets loop back directly ($\overline{w}_1$ and $v$, respectively). □

**Example 5.3.** From the 1-chart $\underline{\mathcal{C}}$ in Ex. 5.1 the 1-chart $\underline{\mathcal{C}}_1$ below arises by eliminating the 1-bisimilarity redundancy $\langle w_1, w_2 \rangle$ through redirecting both of the incoming transitions at $w_1$ over to $w_2$:

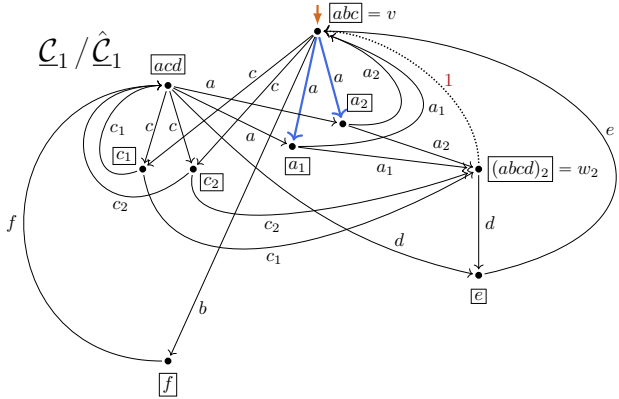

We kept in color only those loop-entry transitions of $\underline{\mathcal{C}}$ that are loop-entry transitions still in $\underline{\mathcal{C}}_1$. The 1-chart $\underline{\mathcal{C}}_1$ is collapsed, because no two different vertices of $\underline{\mathcal{C}}_1$ are 1-bisimilar. But $\underline{\mathcal{C}}_1$ is not a 1-transition free collapse of $\underline{\mathcal{C}}$.

The induced chart $\mathcal{C}_{10} := (\underline{\mathcal{C}}_1)_{(\cdot]}$ of $\underline{\mathcal{C}}_1$ looks as follows:

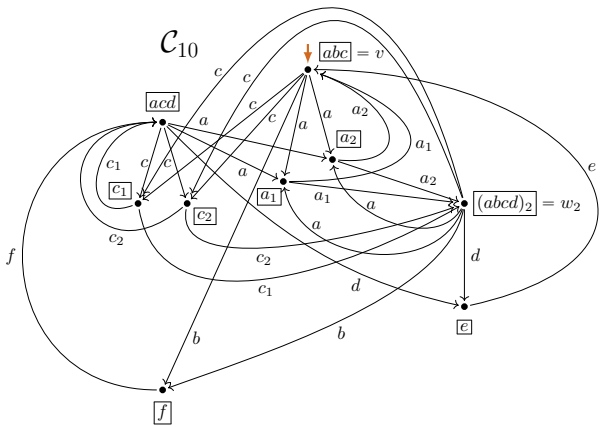

$\mathcal{C}_{10}$ is the 1-transition free bisimulation collapse of $\underline{\mathcal{C}}$.

**Lemma 5.4.** *Neither of the 1-charts $\underline{\mathcal{C}}_1$ and $\mathcal{C}_{10}$ in Ex. 5.3 is a LLEE-1-chart. But both of these 1-charts are collapsed.*

*Proof.* We have already noticed in Ex. 5.3 that $\underline{\mathcal{C}}_1$ and $\mathcal{C}_{10}$ are collapsed. We need to show that neither of them is a LLEE-1-chart.

The 1-chart $\underline{\mathcal{C}}_1$ only contains two transitions that induce loop subcharts: those that are marked in blue. For every other transition $\tau = \langle u_1, \pmb{a}, u_2 \rangle$ of $\underline{\mathcal{C}}_1$ that is not contained in one of these induced loop subcharts it is easy to check that $\tau$ does not induce a loop subchart, because there is an infinite path from $u_1$ starting along $\tau$ that avoids the two blue marked loop-entry transitions, and does not return to the path's source vertex $u_1$. It follows that $\underline{\mathcal{C}}_1$ does not satisfy LEE, the loop existence and elimination condition: after eliminating the two blue marked loop-entry transitions, no other loop-entry transitions exist, yet infinite paths are still possible. Since LLEE implies LEE by Prop. 4.4, it follows that $\underline{\mathcal{C}}_1$ cannot be a LLEE-1-chart.

The chart $\mathcal{C}_{10}$ does not contain a loop-entry transition, which can be checked separately for each transition. Since it expresses infinite behavior, but does not satisfy LEE, $\mathcal{C}_{10}$ is not a LLEE-1-chart. □

To show that neither $\underline{\mathcal{C}}_1$ nor $\mathcal{C}_{10}$ above can be refined into a LLEE-1-chart, we will use the lemma below.

**Lemma 5.5.** *For every finite 1-chart $\underline{\mathcal{D}}$ with LLEE–1-lim there is a finite 1-chart $\underline{\mathcal{D}}'$ with the same induced chart, and with a 1-transition limited LLEE-witness $\hat{\underline{\mathcal{D}}}'$ such that:*

(ptt) *If for a loop-entry identifier $\langle v, n \rangle \in E(\hat{\underline{\mathcal{D}}}')$ the loop sub-1-chart $\mathcal{L}_{\hat{\underline{\mathcal{D}}}'}(v, n)$ of $\underline{\mathcal{D}}'$ at $\langle v, n \rangle$ satisfies both of the following two conditions:*

  *(a) it consists of a single scc (strongly connected component),*
  
  *(b) it contains no other loop vertex than $v$ (that is, $\mathcal{L}_{\hat{\underline{\mathcal{D}}}'}(v, n)$ is an innermost loop sub-1-chart),*

  *then it contains only proper-transition targets in $\underline{\mathcal{D}}'$.*

*Proof idea.* Suppose that $\mathcal{L}_{\hat{\underline{\mathcal{D}}}}(v, n)$ is an innermost loop sub-1-chart of a 1-chart $\underline{\mathcal{D}}$ with 1-transition limited LLEE-witness $\hat{\underline{\mathcal{D}}}$ such that $\mathcal{L}_{\hat{\underline{\mathcal{D}}}}(v, n)$ consists of a single scc, and $v$ is not a proper-transition target. Then $\mathcal{L}_{\hat{\underline{\mathcal{D}}}}(v, n)$ can be shown to contain a vertex $w$ that is a proper-transition target, has a 1-transition back-link to $v$ as its only outgoing transition, and therefore is 1-bisimilar to $v$. See the following example:

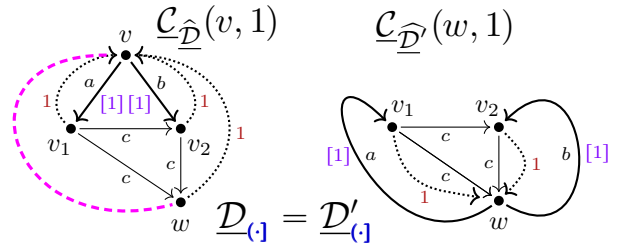

Here we indicate 1-bisimilarity of $v$ and $w$ as magenta link. Then by removing $v$, and letting $w$ take over the role of $v$ as innermost loop vertex, $\mathcal{D}$ can be transformed, preserving induced transitions, into a 1-chart $\mathcal{D}'$ with 1-transition limited LLEE-witness $\hat{\mathcal{D}}'$, in which the innermost loop sub-1-chart $\mathcal{C}_{\hat{\mathcal{D}}'}(v,n)$ now only consists of proper-transition targets.

In general, the start vertex $v$ of $\mathcal{C}_{\hat{\mathcal{D}}}(v,n)$ may have, in $\mathcal{D}$, other departing transitions, and incoming 1-transitions from other loop sub-1-charts that are generated at $v$. In the transformation, all these other outgoing and incoming transitions at $v$ are transferred over to $w$ as well. □

Lem. 5.5 does not hold without the restrictions (a), (b) in (ptt). The 1-charts $\mathcal{D}_1$ and $\mathcal{D}_2$ below with 1-transition limited LLEE-witnesses $\hat{\mathcal{D}}_1$ and $\hat{\mathcal{D}}_2$ show that for (a) and (b).

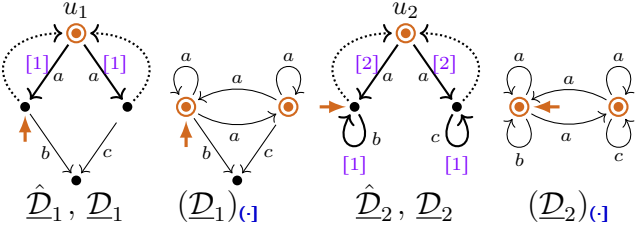

$\hat{\mathcal{D}}_1, \mathcal{D}_1 \qquad (\mathcal{D}_1)_{(\cdot]} \qquad \hat{\mathcal{D}}_2, \mathcal{D}_2 \qquad (\mathcal{D}_2)_{(\cdot]}$

While the loop sub-1-chart at $u_1$ in $\hat{\mathcal{D}}_1$ is not an scc, the loop sub-1-chart at $u_2$ in $\hat{\mathcal{D}}_2$ is not innermost. The induced charts of $\mathcal{D}_1$ and $\mathcal{D}_2$ do not satisfy LEE, and cannot be refined into LLEE-1-charts that only contain proper-transition targets.

**Lemma 5.6.** *Neither of the 1-charts $\mathcal{C}_1$ and $\mathcal{C}_{10}$ in Ex. 5.1 can be refined into a* wg-LLEE-1-chart.

*Proof.* We show the statement of the lemma for $\mathcal{C}_{10}$ by an argument that will demonstrate it also for $\mathcal{C}_1$.

We argue indirectly, towards a contradiction. Suppose that $\mathcal{C}_{10}$ can be refined into a wg-LLEE-1-chart. Then $\mathcal{C}_{10}$ can also be refined, due to Lem. 4.12, into a 1-chart with LLEE–1-lim. Furthermore by Lem. 5.5 it follows, in view of Lem. 3.15, (iv) and (v), that $\mathcal{C}_{10}$ can even be refined into a 1-chart with LLEE–1-lim and (ptt).

Therefore there is a 1-chart $\mathcal{D}$ with 1-transition limited LLEE-witness $\hat{\mathcal{D}}$ and (ptt) such that $\mathcal{D} \to^*_{(1)} \mathcal{C}_{10}$. Since $\mathcal{C}_{10}$ is not a LLEE-1-chart by Lem. 5.4, it follows that $\mathcal{D} \neq \mathcal{C}_{10}$, and therefore also that $\mathcal{D} \to^+_{(1)} \mathcal{C}_{10}$.

We now let $\mathcal{D}_1$ be the 1-chart that results from $\mathcal{D}$ by eliminating, via $\to_{(1)}$ steps, all 1-transition back-links in loop sub-1-charts as delimited by $\hat{\mathcal{D}}$ that are *not* innermost. Then $\mathcal{D} \to^*_{(1)} \mathcal{D}_1 \to^+_{(1)} \mathcal{C}_{10}$ follows by Lem. 3.15, (vi), yielding $\mathcal{D}_1 \to^*_{(1)} \mathcal{C}_{10}$, and where $\mathcal{D}_1 \to^+_{(1)} \mathcal{C}_{10}$ holds, because the innermost loop sub-1-charts in $\mathcal{D}$ must still be present in $\mathcal{D}_1$, but $\mathcal{C}_{10}$ does not contain any loop sub-1-charts. We will show, for a 1-chart $\mathcal{C}_2$ that will be illustrated below and is similar to $\mathcal{C}_1$, the following contradictory statements:

(S1) $\mathcal{D} \to^+_{(1)} \mathcal{D}_1 \to_{(1)} \mathcal{C}_{10}$ for $\mathcal{D}_1 \in \{\mathcal{C}_1, \mathcal{C}_2\}$, with $\mathcal{D} \neq \mathcal{D}_1$.
(S2) Neither $\mathcal{C}_1$ nor $\mathcal{C}_2$ can be refined into a 1-chart with LLEE–1-lim.

The contradiction arises between (S2) and the part of (S1) that states that either $\mathcal{C}_1$ or $\mathcal{C}_2$ can be refined into $\mathcal{D}$ which we assumed is a 1-chart with LLEE–1-lim. Therefore it remains to define $\mathcal{C}_2$ and to prove (S1) and (S2).

Turning to defining $\mathcal{C}_2$ and showing (S1), we use that $\mathcal{D}$ and $\hat{\mathcal{D}}$ satisfy (ptt). Therefore all innermost loop sub-1-charts of $\mathcal{D}$ delimited by $\hat{\mathcal{D}}$ consist only of proper-transition targets in $\mathcal{D}$. Since back-links in these innermost loop sub-1-charts of $\mathcal{D}$ are only removed in the steps $\mathcal{D}_1 \to^+_{(1)} \mathcal{C}_{10}$, it follows that $\mathcal{D}_1$ has the same vertices as $\mathcal{C}_{10}$. Conversely, these innermost loop sub-1-charts of $\mathcal{D}$ can be made visible starting from $\mathcal{C}_{10}$ by adding 1-transition back-links between existing vertices of $\mathcal{C}_{10}$, and by removing transitions that become superfluous in the sense that they correspond to newly formed induced transitions. However, to obtain a 1-bisimilar refinement any such 1-transition back-link $\langle u, 1, u' \rangle$ can only be added to $\mathcal{C}_{10}$ if $u'$ is a substate of $u$ already in $\mathcal{C}_{10}$, and that is, if $u'$ has part of the behavior of $u$. There are only two possibilities for such an added back-link: one from $w_2$ to $v$ (due to $v \sqsubseteq_{\mathcal{C}_{10}} w_2$), and another one from $w_2$ to $\overline{w}_1$ (due to $\overline{w}_1 \sqsubseteq_{\mathcal{C}_{10}} w_2$). It is not possible to add them both as 1-transitions, because as such they correspond to back-links in the 1-transition limited LLEE-witness $\hat{\mathcal{D}}$, and no vertex can loop back directly to two different loop vertices in a LLEE-witness. It follows that $\hat{\mathcal{D}}$ can contain only one innermost loop vertex, namely either $v$ or $\overline{w}_1$.

Therefore in order to make visible, starting from $\mathcal{C}_{10}$, the innermost loop subchart of $\mathcal{D}$ defined by $\hat{\mathcal{D}}$, we either obtain the 1-chart $\mathcal{C}_1$ in Ex. 5.3, if $v$ is a loop vertex and $\langle w_2, 1, v \rangle$ is a back-link of $\hat{\mathcal{D}}$, or the 1-chart $\mathcal{C}_2$ below:

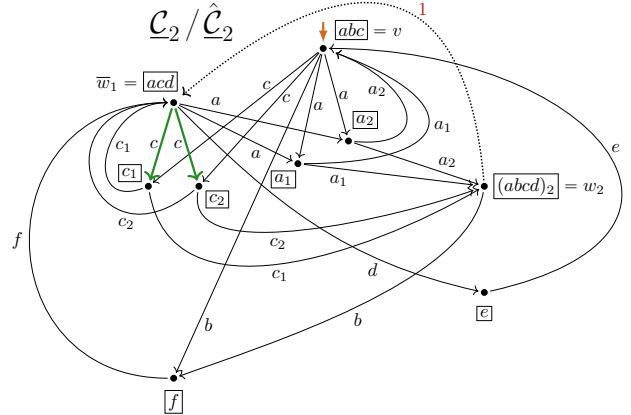

if $\overline{w}_1$ is a loop vertex, and $\langle w_2, 1, \overline{w}_1 \rangle$ is a back-link of $\hat{\mathcal{D}}$. It is straightforward to check that only the two transitions marked in green induce loop sub-1-charts of $\mathcal{C}_2$, and that after eliminating those, no further loop-entry transitions arise, but infinite paths remain. Therefore $\mathcal{C}_2$ cannot be a LLEE-chart. It follows that $\mathcal{D}$ must be a further refinement either of $\mathcal{C}_1$ or of $\mathcal{C}_2$. In this way we have established (S1).

For (S2) we focus on $\mathcal{C}_2$, because the argument for $\mathcal{C}_1$ is closely analogous. We assume that a LLEE-1-chart $\mathcal{D}$ with 1-transition limited LLEE-witness $\hat{\mathcal{D}}$ is a refinement of $\mathcal{C}_2$. Then $\mathcal{D} \to^+_{(1)} \mathcal{C}_2$ follows, because $\mathcal{C}_2$ is not a LLEE-1-chart.

Now we note that $\mathcal{C}_2$ cannot be refined in the direction of the 1-transition limited LLEE-witness $\hat{\mathcal{D}}$ by adding 1-transition back-links to vertices that are already present in $\mathcal{C}_2$. This is because, to satisfy the layeredness condition (W3) for the LLEE-witness $\hat{\mathcal{D}}$, any such back-link needed to start from $\overline{w}_1$, or from a vertex outside of $\mathcal{C}_{\hat{\mathcal{D}}}(\overline{w}_1, 1)$ (the sub-1-chart of $\mathcal{C}_2$ delimited by $\hat{\mathcal{D}}$ and by its entry/body-labeling $\hat{\underline{\mathcal{C}}}_2$), and target a different of these vertices that furthermore needs to be a substate. But there is not such a vertex in $\mathcal{C}_2$: neither $\overline{w}_1$ nor one of $v$, $\boxed{a_1}$, $\boxed{a_2}$, $\boxed{e}$ outside of $\mathcal{C}_{\hat{\mathcal{D}}}(\overline{w}_1, 1)$ has a substate among the other 4 of these vertices.

Consequently, the only option to refine $\hat{\underline{\mathcal{C}}}_2$ towards the 1-transition limited LLEE-witness $\hat{\mathcal{D}}$ is to add a vertex $u$ that is not a proper-transition target, with 1-transition back-links directed to it from one of the five vertices $\overline{w}_1$, $v$, $\boxed{a_1}$, $\boxed{a_2}$, or $\boxed{e}$ such that proper transitions in $\mathcal{C}_2$ arise as induced transitions via $u$. The only option to simplify the structure is to share, via the new vertex $u$ proper transitions from two of these five vertices. Such a non-trivial sharing of transitions is possible only between $\overline{w}_1$ and $v$, because none of other pairs of vertices have an action label of an outgoing transition in common. The maximal option is to share, via the new vertex $u$, all outgoing $a$- and $c$-transitions from $\overline{w}_1$ and $v$. This leads to the 1-chart $\underline{\mathcal{C}}_2'$:

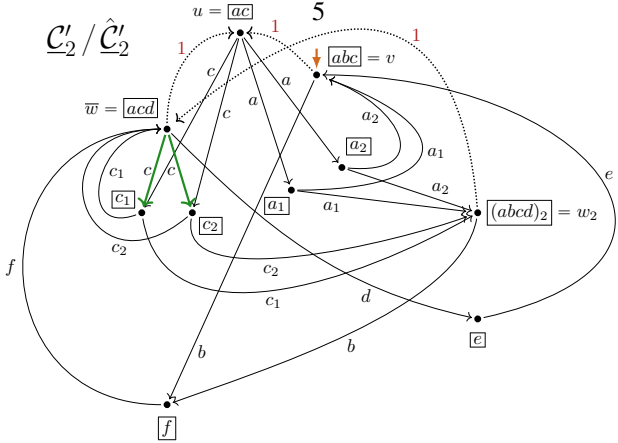

But we note now that no loop-entry transition is created at $u$ in $\underline{\mathcal{C}}_2'$, and so the added 1-transitions are not back-links. So this refinement step is not one in the direction of the (assumed) 1-transition limited LLEE-witness $\hat{\mathcal{D}}$. While it would be possible to share only a non-empty subset of the four transitions from $u$, that would not give rise to a loop-entry transition at $u$, either. Hence we must recognize that it is not possible to refine $\mathcal{C}_2$ into the LLEE-1-chart $\mathcal{D}$ as assumed. Thus we have shown the part of (S2) for $\mathcal{C}_2$.

Analogously it can be argued that neither can $\mathcal{C}_1$ be refined into a LLEE-1-chart, thereby obtaining (S2).

Having obtained a contradiction, we conclude that neither $\mathcal{C}_{10}$ nor $\mathcal{C}_1$ can be refined into a wg-LLEE-1-chart. □

On the basis of these preparations we can now formulate and prove our two main results concerning the non-preservation of two properties under bisimulation collapse.

**Theorem 5.7.** *The property of a 1-chart $\mathcal{D}$ that it can be refined by a wg-LLEE-1-chart is not preserved in general under the step from $\mathcal{D}$ to its bisimulation collapse.*

*Proof.* The LLEE-1-chart $\mathcal{D} := \mathcal{C}$ in Ex. 5.1 together with its (1-transition free) bisimulation collapse $\mathcal{C}_{10}$ in Ex. 5.3 provides a counterexample, because $\mathcal{C}_{10}$ cannot be refined into a wg-LLEE-1-chart due to Lem. 5.6. □

**Theorem 5.8.** *The image of the chart interpretation $\mathcal{C}(\cdot)$ is not closed, up to isomorphism, under bisimulation collapse.*

*Proof.* A counterexample consists of the induced chart $\underline{\mathcal{C}}_{(\cdot)}$ of $\mathcal{C}$ in Ex. 5.1, and its bisimulation collapse $\mathcal{C}_{10}$ in Ex. 5.3.

The induced chart $\underline{\mathcal{C}}_{(\cdot)}$ of $\underline{\mathcal{C}}$ is, up to isomorphism, in the image of $\mathcal{C}(\cdot)$ due to Lem. 5.2.

But its collapse $\mathcal{C}_{10}$ of $\underline{\mathcal{C}}_{(\cdot)}$ is not isomorphic to the chart interpretation of any star expression. In order to show this, we assume that $\mathcal{C}_{10} \simeq \mathcal{C}(e_{10})$ for some star expression $e_{10} \in StExp(A)$, and will derive a contradiction. By Cor. 3.16 we find that $\mathcal{C}(e_{10})$ can be refined into the weakly guarded 1-chart interpretation $\underline{\mathcal{C}}(e_{10})$ of $e_{10}$, where $\underline{\mathcal{C}}(e_{10})$ is a LLEE-1-chart by Thm. 4.7. Therefore $\mathcal{C}(e_{10})$ can be refined into a wg-LLEE-1-chart. Now the property of a 1-chart to be refinable by 1-transitions into a wg-LLEE-1-chart is invariant under isomorphism. Therefore the assumption $\mathcal{C}_{10} \simeq \mathcal{C}(e_{10})$ implies that $\mathcal{C}_{10}$ can be refined into a wg-LLEE-1-chart. But this is a contradiction with the statement of Lem. 5.6. □

**The origin of the counterexample.** A procedure for constructing the bisimulation collapse of a finite LLEE-chart by LLEE-preserving steps that collapse pairs of bisimilar vertices was described in [10]. This procedure repeatedly removes, from a given LLEE-chart, one vertex $w_1$ from a pair $\langle w_1, w_2 \rangle$ of distinct, bisimilar vertices, after redirecting all incoming transitions at $w_1$ over to $w_2$, thereby obtaining a bisimilar chart. Although only vertices from 'reduced' pairs in specified positions can be eliminated LLEE-preservingly, the bisimulation collapse is reached, because every LLEE-chart that is not collapsed contains a reduced pair.

Pursuing the aim of generalizing that stepwise collapse procedure to LLEE-1-charts, we have explored how position conditions for 'reduced' pairs of 1-bisimilar vertices can be defined for 1-charts with LLEE–1-lim. This led us to position conditions where among four that generalize condition (C3) in [10] for reduced pairs $\langle w_1, w_2 \rangle$, where $w_1$ and $w_2$ with 1-bisimilar and in the same scc, are the following two:

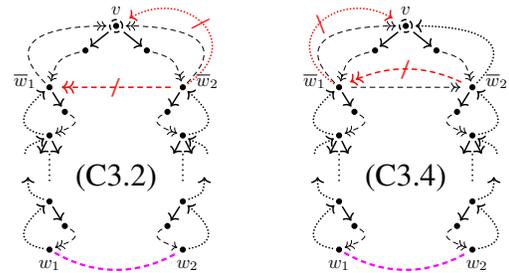

In both of these cases the 1-bisimilar vertices $w_1$ and $w_2$ are in nested loop sub-1-charts below different vertices $\overline{w}_1$ and $\overline{w}_2$, respectively, where $\overline{w}_1$ and $\overline{w}_2$ are in the body of a sub-1-chart of a loop vertex $v$, and $w_1$ and $w_2$ are linked by 1-transition back-links to $\overline{w}_1$ and $\overline{w}_2$, respectively.

Moreover, $\overline{w}_2$ is not linked to $\overline{w}_1$ via body transitions only. However, $\overline{w}_1$ can be reached from $\overline{w}_2$ via $v$ and a loop-entry transition from $v$. In (C3.2) a 1-transition back-link is excluded from $\overline{w}_2$ to $v$, and in (C3.4) from $\overline{w}_1$ to $v$.

While we succeeded in finding LLEE-preserving transformations that eliminate $w_1$ in case (C3.2) (see Ex. 6.1) and in 4 other cases, we did not find such a transformation for case (C3.4). That failure led us, after narrowing down the hard problem, to the construction of the counterexample LLEE-1-chart $\underline{\mathcal{C}}$ in Ex. 5.1, which is indeed of form (C3.4).

## 6. Conclusion, and outlook

In order to approach the axiomatization and expressibility problems for Milner's process semantics of regular expressions we undertook two preparatory steps: a conceptual one, and one that highlights a crucial difficulty. First, we showed that process interpretations of regular expression can be refined by 1-transitions (Cor. 3.16) into graphs that satisfy LEE (Thm. 4.7). This clears the way for generalizing the methods used for results about 1-free regular expressions in [10]. But second, we discovered that the refinement property is not preserved for process interpretations under bisimulation collapse (Thm. 5.7). This phenomenon sets the problems **(E)** and **(A)** for the full class of regular expressions apart from their specialization to subclasses with restricted uses of 0 and/or 1, such as the 1-free expressions in [10].

In order to make this observation, we found it crucial (i) to view process interpretations of regular expressions as graphs (possibly with 1-transitions) that are constrained by the loop existence and elimination property LEE (from [10]), and (ii) to work with layered LEE-witnesses that represent process interpretations of regular expressions (at least) up to bisimilarity (a correspondence established in [10]).

As a consequence of the counterexample there cannot be a LLEE-preserving procedure for producing the bisimulation collapse of LLEE-1-charts. But we found the following positive sign: stepwise elimination of a vertex from a reduced pair of 1-bisimilar vertices is almost always possible, with the exception of pairs of the form (C3.4) above.

**Example 6.1.** In the LLEE-1-chart $\underline{\mathcal{D}}$ below we assume all actions are the same, and indicate a LLEE-witness by green loop-entry transitions of level 1, and blue ones of level 2. $\underline{\mathcal{D}}$ contains the pair $\langle w_1, w_2 \rangle$ of 1-bisimilar vertices that is of 'reduced' form (C3.2). Here $w_1$ can be eliminated by three steps two of which redirect transitions to 1-bisimilar targets:

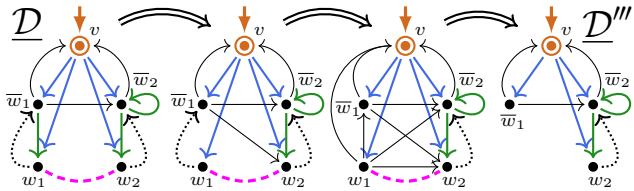

The first step removes the loop at $\overline{w}_1$, the second restores LLEE–1-lim, and in the third $w_1$ is eliminated. The result is the 1-chart $\underline{\mathcal{D}}'''$ with LLEE–1-lim that is 1-bisimilar to $\underline{\mathcal{D}}$.

A procedure that eliminates all reduced 1-bisimilarity redundancies except those of form (C3.4) is able to approximate the bisimulation collapse of LLEE-1-charts while preserving LLEE. Additional simplifications suggest that the number of distinct, 1-bisimilar vertices in a minimized LLEE-1-chart can be reduced to two, like in $\underline{\mathcal{C}}$ in Ex. 5.1. We believe that such an approximation procedure can be a crucial step in tackling the general problems **(E)** and **(A)**.

**Relation to subsequent work that was published earlier.** This article was written as a preparation for developing a solution of Milner's axiomatization problem **(A)**. We did so by explaining why the proof-strategy that Fokkink and I had employed in [10] for solving the special case of problem **(A)** for 1-free star expressions in [10] does not generalize, at least not directly, to a successful proof strategy for the problem **(A)** itself. Indeed, the proof-strategy in [10] was based on the property **(C)** (preservation of LEE under bisimulation collapse for charts). But we demonstrated here that **(C)** does not generalize to 1-charts, not even in its refinement form **(RC)** (preservation under bisimulation collapse of the refinability of a 1-chart into a LLEE-1-chart). While this failure (RC) of generalizing **(C)** to LLEE-1-charts only makes it impossible to use **(RC)** for solving problem **(A)**, we expected that this fact forms a conceptual complication and significant stumbling block for *any* completeness proof of Milner's proof system. At the same time we already had in mind a possible solution to navigate this difficulty: the use of LLEE-1-chart approximations of bisimulation collapses.

Indeed, we eventually described such a solution for **(A)** later in the summary [15] for LICS 2022 of a completeness proof of Milner's system. There, we based the development on the counterexample for LLEE-preserving bisimulation collapse that we described here in Section 5. Importantly, we abstracted its main constitutive features in order to define the concept of 'twin-crystal', for which the 1-chart $\underline{\mathcal{C}}$ in Ex. 5.1 is a prime example. Then we crucially defined a 'crystallization' technique for the minimization of LLEE-1-charts under bisimilarity with as results LLEE-1-charts that tightly approximate bisimulations collapses. We show that every weakly guarded LLEE-1-chart can be minimized under bisimilarity to a 'crystallized' LLEE-1-chart that is collapsed with the possible exception of some strongly connected components that are of twin-crystal form.

Before we wrote down the solution of **(A)** using the crystallization technique (in autumn 2021), we set out (in spring 2021) to find some concrete evidence for our expectation as expressed above: that *any* completeness proof of Milner's proof system Mil has to navigate, in some way or the other, the difficulty that is presented by the failure (RC) of LLEE-preserving bisimulation collapse of LLEE-1-charts. We sought to obtain such evidence by investigating, and trying to characterize, the derivational power of the *single-equation fixed-point rule* RSP* in Milner's system.[1] Indeed it is the a priori limited derivational power of that rule that presents the main difficulty for a completeness proof of Mil. This is because the system that arises from Mil by adopting a

---

[1]. RSP* means 'recursive specification principle' for single-fixed point equations by using iteration. From the fixed-point equation $e = f \cdot e + g$ it permits to infer the specification $e = f^* \cdot g$ of $e$, provided that $f \!\Downarrow$ holds.

*fixed-point rule* USP *for guarded systems of equations* (here USP stands short for 'unique solvability principle') can be shown to be complete in a rather straightforward way.

In the article [16] for CALCO 2021 with report [17] we showed that the derivational power of the single-equation fixed-point rule RSP* in Mil corresponds to the derivational power of a coinductive rule. That rule permits cyclic derivations of the form of guarded LLEE-1-charts in order to infer that two Mil-provable solutions (which occur in the derivations) of the used LLEE-1-chart are equal (at the start vertex of the LLEE-1-chart). By adding this coinductive rule to Mil and deleting the single-equation fixed-point rule RSP*, we defined a coinductive version cMil of Milner's system Mil. Then we showed by proof-theoretical interpretations in both directions that cMil and Mil are theorem-equivalent.

In [18] we extended the exhibition of this result in several ways. We also argued there that, since the coinductive version cMil of Mil can be viewed as being situated roughly half-way in between Mil and bisimulations between chart interpretations of star expressions, this new system cMil may become a natural beachhead for a completeness proof of Mil.

While we did not directly use the coinductive system cMil as a beachhead for the completeness proof later in [15], the characterization of derivability in Mil by derivability in cMil in [16] convinced us of the expediency of the underlying concepts. In particular, this result had tied derivability in Mil closely to the concept of guarded LLEE-1-charts for expressing process interpretations of regular expressions. Use of this concept, together with the fact that LLEE-1-charts are uniquely solvable modulo provability in Mil (as shown in [16], [17], [18]) were crucial stepping stones for the completeness proof of Mil that we summarize in [15].

## Acknowledgment

I want to thank a reviewer for LICS 2021 for (i) pointing out some deficient details (which are corrected here), and (ii) for mentioning in the review that the sticking point for the submission in the PC discussion was that "it is indeed only showing that a specific approach towards tackling **(A)** is bound to fail" (our intuitive disagreement with this judgment motivated our work in [16], [17] where we linked LLEE-1-charts closely to proofs in Milner's system, see above).

# Appendix A.
# Supplements, more proof details, and omitted proofs

## A.1. Supplements for Section 2: Process semantics of star expressions

**Lemma** (= Lem. 2.5). *For every star expression $e \in StExp(A)$, the chart interpretation $\mathcal{C}(e)$ is a finite chart.*

*Proof (method).* Derivatives of a star expressions with respect to the simpler TSS $\mathcal{T}_0(A)$ in Rem. 2.6 coincide with Antimirov's partial derivatives. Then finiteness of chart interpretations defined according to $\mathcal{T}_0(A)$ follows from Antimirov's result [2] that every regular expression only has finitely many iterated derivatives. In order to prove finiteness of chart interpretations as defined via $\mathcal{T}(A)$ in Def. 3.6, a correspondence with a variation of partial derivatives can be defined, and for that a finiteness results can be shown analogously. □

**Lemma A.1.** *The following statements hold concerning 1-free and under-star-1-free star expressions, and concerning their derivatives and iterated derivatives:*

(i) $StExp^{(\mathbf{1})}(A) \subseteq StExp^{(*/\mathbf{1})}(A)$ *holds for all sets $A$ of actions, that is, every 1-free star expression is also an under-star-1-free star expression.*

(ii) *Derivatives and iterated derivatives of 1-free star expressions are under-star-1-free star expressions that are generated by the following grammar:*

$$fd ::= 1 \mid fd \cdot f \mid (fd \cdot (f_1)^*) \cdot f_2 \quad (\text{where } f, f_1, f_2 \in StExp^{(\mathbf{1})}(A))$$

(iii) *Derivatives and iterated derivatives of under-star-1-free star expressions are under-star-1-free star expressions that are generated by the following grammar:*

$$ufd ::= 1 \mid ufd \cdot uf \quad (\text{where } uf \in StExp^{(*/\mathbf{1})}(A))$$

*It follows that the class of under-star-1-free star expressions (and the sets $StExp^{(*/\mathbf{1})}(A)$ for all sets $A$) are closed under taking derivatives.*

**Lemma** (= Lem. 2.11). *Every 1-free star expression is also an under-star-1-free star expression, and hence $StExp^{(\mathbf{1})}(A) \subseteq StExp^{(*/\mathbf{1})}(A)$. The set $StExp^{(*/\mathbf{1})}(A)$ of under-star-1-free star expressions over $A$ is closed under taking derivatives.*

*Proof.* $StExp^{(\mathbf{1})}(A) \subseteq StExp^{(*/\mathbf{1})}(A)$ is the statement of Lem. A.1, (i). Closedness of $StExp^{(*/\mathbf{1})}(A)$ under taking derivatives follows from Lem. A.1, (iii). □

**Proposition** (= Prop. 2.12). *The image of the class $StExp^{(*/\mathbf{1})}(A)$ of under-star-1-free star expressions under the chart interpretation $\mathcal{C}(\cdot)$ is closed under the operation of bisimulation collapse, modulo isomorphism.*

This statement can be shown (see on page 18 below) with a refinement of the results and the methods for the class $StExp_{\mathbf{1}}^{\circledast}(A)$ of 1-free star expressions with binary star iteration $\circledast$ in [10], [14]. Here we base ourselves on the class $StExp^{(*/\mathbf{1})}$ of under-star-1-free star expressions that extends, in a natural way, the class $StExp^{(\mathbf{1})}$ of 1-free star expressions with unary star iteration as considered here that corresponds to the class $StExp_{\mathbf{1}}^{\circledast}(A)$ of star expressions with binary star iteration in [10], [14]. While Prop. 2.12 also holds for the subclass $StExp^{(\mathbf{1})}$ of 1-free star expressions within the class $StExp^{(*/\mathbf{1})}$ of under-star-1-free star expressions, we formulate and prove it only for the larger class here. For being able to prove Prop. 2.12, the more careful formulation of the TSS in Def. 2.4 is crucial, see Rem. 2.6, which uses normedness case distinctions in the rules for product and star.

In [10] the first two of the following statements have been shown; the third is a property of the extraction procedure for provable solutions defined there:

**(I)$_{\mathbf{1}}^{\circledast}$** The chart interpretation $\mathcal{C}(e)$ of a 1-free star expression $e \in StExp_{\mathbf{1}}^{\circledast}(A)$ is a LLEE-chart. (See Proposition 3.7 in [10]).

**(C)** The bisimulation collapse of a LLEE-chart is again a LLEE-chart. (See Theorem 6.9 in [10]).

**(E)$_{\mathbf{1}}^{\circledast}$** From every LLEE-chart $\mathcal{C} = \langle V, A, v_s, \rightarrow, \downarrow \rangle$ with LLEE-witness $\hat{\mathcal{C}}$ a function $s_{\hat{\mathcal{C}}} : V \rightarrow StExp_{\mathbf{1}}^{\circledast}(A)$ that produces 1-free star expressions can be extracted such that $\mathcal{C} \Leftrightarrow \mathcal{C}(s_{\hat{\mathcal{C}}}(v_s))$ holds, that is, such that $\mathcal{C}$ is bisimilar to the chart interpretation $\mathcal{C}(s_{\hat{\mathcal{C}}}(v_s))$ of $s_{\hat{\mathcal{C}}}(v_s)$. (This can be shown for the extraction function defined in Definition 5.3 in [10] where $s_{\hat{\mathcal{C}}}$ is shown to be a solution of $\mathcal{C}$ that is provable in the proof system **BBP**, see Proposition 5.5 in [10].)

Statement **(E)$_{\mathbf{1}}^{\circledast}$** can be established by showing that the relation $graph(s_{\hat{\mathcal{C}}}) \cdot \left( \Leftrightarrow_{\mathcal{L}(StExp(A))} \cap \left( V(\mathcal{C}(s_{\hat{\mathcal{C}}}(v_s))) \times V(\mathcal{C}(s_{\hat{\mathcal{C}}}(v_s))) \right) \right)$, the composition of the graph of $s_{\hat{\mathcal{C}}}$ with bisimilarity $\Leftrightarrow_{\mathcal{L}(StExp(A))}$ on the LTS defined by the process (chart) interpretation of all star expressions but restricted to $\mathcal{C}(s_{\hat{\mathcal{C}}}(v_s))$, is a bisimulation between $\mathcal{C}$ and $\mathcal{C}(s_{\hat{\mathcal{C}}}(v_s))$.

Now **(I)$_{\mathbf{1}}^{\circledast}$** can be transferred in a straightforward manner to apply to under-star-1-free star expressions in $StExp^{(*/\mathbf{1})}$ (see Def. 2.1, and thus with unary star). Furthermore, by refining the extraction procedure, and thus by carefully redefining solutions extracted from LLEE-charts, statement $\mathbf{(E)}_{\mathbf{1}}^{\circledast}$ can be strengthened to stating that there is actually a functional bisimulation from a LLEE-chart $\mathcal{C}$ to the chart interpretation $\mathcal{C}(s_{\hat{\mathcal{C}}}(v_{\text{s}}))$ of the solution extracted from $\hat{\mathcal{C}}$. We gather these two statements:

$\mathbf{(I)_1}$ The chart interpretation $\mathcal{C}(e)$ of a under-star-1-free star expression $e \in StExp^{(*/\mathbf{1})}(A)$ is a LLEE-chart.

$\mathbf{(E)}_{\mathbf{1}}^{\mathbf{(ref)}}$ From every LLEE-chart $\mathcal{C} = \langle V, A, v_{\text{s}}, \rightarrow, \downarrow \rangle$ with LLEE-witness $\hat{\mathcal{C}}$ a $\mathsf{Mil}^-$-provable solution $s_{\hat{\mathcal{C}}} : V \to StExp^{(*/\mathbf{1})}(A)$ of $\mathcal{C}$ can be extracted that has the property that $\mathcal{C} \rightrightarrows_{s_{\hat{\mathcal{C}}}} \mathcal{C}(s_{\hat{\mathcal{C}}}(v_{\text{s}}))$ holds, that is, $s_{\hat{\mathcal{C}}}(v_{\text{s}})$ also defines, as its graph, a functional bisimulation from $\mathcal{C}$ to the chart interpretation $\mathcal{C}(s_{\hat{\mathcal{C}}}(v_{\text{s}}))$ of $s_{\hat{\mathcal{C}}}(v_{\text{s}})$. (See Lem. A.10 below.)

Now we describe the adaptation of the extraction procedure that is necessary to prove $\mathbf{(E)}_{\mathbf{1}}^{\mathbf{(ref)}}$. In order to bring solutions extracted from LLEE-witnesses into close correspondence with derivatives as defined by the TSS $\mathcal{T}$, the redefinition of extracted solutions needs to refine the definition of extracted solutions in Definition 5.3 in [10] in the following two respects:

(a) An 'initialized' variant $T_{\hat{\mathcal{C}}}(e, w, v)$ of the relative extraction function $t_{\hat{\mathcal{C}}}(w, v)$ is defined, for $w, v \in V$ with $w \curvearrowright v$ and an 'initializing' star expression $e$, directly only in the two cases $w \, {}_d\!\!\circlearrowright v$ (that is, $w$ directly loops back to $v$), and $w \not\circlearrowright$ (that is, $w$ does not loop back to another vertex). Additionally it is extended by an inductive clause to the remaining case $w \, {}_d\!\!\circlearrowright \overline{w} \neq v$, (that is, $w$ loops back directly to another vertex $\overline{w}$ different from $v$); then $w$ loops back directly $w \, {}_d\!\!\circlearrowright \overline{w} = \overline{w}_1 \, {}_d\!\!\circlearrowright \ldots \, {}_d\!\!\circlearrowright \overline{w}_l$ to $\overline{w} = \overline{w}_1$ and further vertices $\overline{w}_2, \ldots, \overline{w}_l$ with $l \geq 2$, thereby either eventually reaching with $v = w_l$, or not reaching $v$ at all in case that $v \notin \{w_1, \ldots, w_l\}$ and $\overline{w}_l \not\circlearrowright$ hold.

(b) In order to take into account that derivatives of under-star-1-free star expressions always create an occurrence of 1 at the left of the leftmost product (concatenation) occurrence in the expression, an 'initialized' version $T_{\hat{\mathcal{C}}}(e, w, v)$ of the relativized extraction function has to be defined that puts a star expression $e$ immediately left of the leftmost product position of the (as in (a)) adapted relativized extraction function $t_{\hat{\mathcal{C}}}$ that we will call initialized and denote by $T_{\hat{\mathcal{C}}}$.

**Notation A.2.** For a LLEE-witness $\hat{\mathcal{C}} = \langle V, A, v_{\text{s}}, \hat{\rightarrow}, \downarrow \rangle$ of a chart $\mathcal{C} = \langle V, A, v_{\text{s}}, \rightarrow, \downarrow \rangle$ we introduce the following notation for the set of transitions that depart from a given vertex $w \in V$:

$$\{w \hat{\rightarrow}\} := \hat{\rightarrow} \cap \big(\{w\} \times (A \times \mathbb{N}) \times V\big)$$
$$= \{\langle w, \langle a, l \rangle, w' \rangle \mid \langle w, \langle a, l \rangle, w' \rangle \in \hat{\rightarrow}\}.$$

**Definition A.3.** Let $\mathcal{C} = \langle V, A, v_{\text{s}}, \rightarrow, \downarrow \rangle$ be a LLEE-chart with LLEE-witness $\hat{\mathcal{C}} = \langle V, A, v_{\text{s}}, \hat{\rightarrow}, \downarrow \rangle$. The *initialized relative extraction function* $T_{\hat{\mathcal{C}}}(\cdot, \cdot, \cdot)$ of $\hat{\mathcal{C}}$ is the following function with star expression values:

$$T_{\hat{\mathcal{C}}}(\cdot, \cdot, \cdot) : StExp(A) \times ((V \times V) \cap \curvearrowleft) \longrightarrow StExp(A)$$
$$\langle e, w, v \rangle \longmapsto T_{\hat{\mathcal{C}}}(e, w, v),$$

that are defined, for all *initializing star expressions* $e \in StExp(A)$, and for all $w, v \in V$ with $w \curvearrowleft v$ as follows:

$$T_{\hat{\mathcal{C}}}(e, w, v) := \begin{cases} \big(e \cdot \big(\sum_{i=1}^{n_1} a_i + \sum_{i=1}^{n_2} T_{\hat{\mathcal{C}}}(b_i, w_i, w)\big)^*\big) \cdot \big(\sum_{i=1}^{n_4} T_{\hat{\mathcal{C}}}(d_i, u_i, v)\big) & \text{if } w \not\circlearrowright \text{ and (A.1)}, \\ \big(e \cdot \big(\sum_{i=1}^{n_1} a_i + \sum_{i=1}^{n_2} T_{\hat{\mathcal{C}}}(b_i, w_i, w)\big)^*\big) \cdot \big(\sum_{i=1}^{n_3} c_i + \sum_{i=1}^{n_4} T_{\hat{\mathcal{C}}}(d_i, u_i, v)\big) & \text{if } w \, {}_d\!\!\circlearrowright v \text{ and (A.2)}, \\ T_{\hat{\mathcal{C}}}(T_{\hat{\mathcal{C}}}(e, w, \overline{w}), \overline{w}, v) & \text{if } w \, {}_d\!\!\circlearrowright \overline{w} \neq v, \end{cases}$$

where we assume that the set of transitions from $w$ in $\hat{\mathcal{C}}$ have the representations:

$$\{w \hat{\rightarrow}\} = \{w \xrightarrow{a_1}_{[l_1]} w, \ldots, w \xrightarrow{a_{n_1}}_{[l_{n_1}]} w\} \cup \{w \xrightarrow{b_1}_{[l'_1]} w_1, \ldots, w \xrightarrow{b_{n_2}}_{[l'_{n_2}]} w_{n_2}\}$$
$$\cup \{w \xrightarrow{d_1}_{\text{bo}} u_1, \ldots, w \xrightarrow{d_{n_4}}_{\text{bo}} u_{n_4}\} \quad \text{(A.1)}$$
$$\text{with } w_1, \ldots, w_{n_2} \neq w,$$

$$\{w \hat{\rightarrow}\} = \{w \xrightarrow{a_1}_{[l_1]} w, \ldots, w \xrightarrow{a_{n_1}}_{[l_{n_1}]} w\} \cup \{w \xrightarrow{b_1}_{[l'_1]} w_1, \ldots, w \xrightarrow{b_{n_2}}_{[l'_{n_2}]} w_{n_2}\}$$
$$\cup \{w \xrightarrow{c_1}_{\text{bo}} v, \ldots, w \xrightarrow{c_{n_3}}_{\text{bo}} v\} \cup \{w \xrightarrow{d_1}_{\text{bo}} u_1, \ldots, w \xrightarrow{d_{n_4}}_{\text{bo}} u_{n_4}\}, \quad \text{(A.2)}$$
$$\text{with } w_1, \ldots, w_{n_2} \neq w, \text{ and } u_1, \ldots, u_{n_4} \neq v,$$

for $n_1, n_2, n_3, n_4 \in \mathbb{N}$, $a_1, \ldots, a_{n_1}, b_1, \ldots, b_{n_2}, c_1, \ldots, c_{n_3}, d_1, \ldots, d_{n_4} \in A$, and $l_1, \ldots, l_{n_1}, l'_1, \ldots, l'_{n_2} \in \mathbb{N} \setminus \{0\}$, and where the definition proceeds by induction on the lexicographic ordering $<_{\text{lex}}$ of the well-founded relations $\leftarrow_{\text{bo}}^+$ and $\curvearrowleft^+$ on pairs $\langle w, v \rangle$ of vertices with precedence given to the second component such that $<_{\text{lex}}$ is defined, for all $w_1, w_2, v_1, v_2 \in V$, by:

$$\langle w_1, v_1 \rangle <_{\text{lex}} \langle w_2, v_2 \rangle \;:\Longleftrightarrow\; v_1 \curvearrowleft^+ v_2 \,\vee\, \big(v_1 = v_2 \,\wedge\, w_1 \leftarrow_{\text{bo}}^+ w_2\big). \quad \text{(A.3)}$$

**Definition A.4.** Let $\mathcal{C} = \langle V, A, v_s, \to, \downarrow \rangle$ be a LLEE-chart with LLEE-witness $\hat{\mathcal{C}} = \langle V, A, v_s, \hat\to, \downarrow \rangle$. The extraction function $s_{\hat{\mathcal{C}}}$ of $\hat{\mathcal{C}}$ is defined by using the initialized extraction function $S_{\hat{\mathcal{C}}}(\cdot, \cdot)$ of $\hat{\mathcal{C}}$, where these functions have the types:

$$s_{\hat{\mathcal{C}}} : V \longrightarrow StExp(A) \qquad\qquad S_{\hat{\mathcal{C}}}(\cdot, \cdot) : StExp(A) \times V \longrightarrow StExp(A)$$
$$w \longmapsto s_{\hat{\mathcal{C}}}(w), \qquad\qquad\qquad \langle e, w \rangle \longmapsto S_{\hat{\mathcal{C}}}(e, w),$$

and are defined, for all initializing star expressions $e \in StExp(A)$, and for all $w \in V$, as follows:

$$s_{\hat{\mathcal{C}}}(w) := S_{\hat{\mathcal{C}}}(1, w),$$

$$S_{\hat{\mathcal{C}}}(e, w) := \begin{cases} \left( e \cdot \left( \sum_{i=1}^{n_1} a_i + \sum_{i=1}^{n_2} T_{\hat{\mathcal{C}}}(b_i, w_i, w) \right)^* \right) \cdot \left( \tau_{\mathcal{C}}(w) + \sum_{i=1}^{n_4} S_{\hat{\mathcal{C}}}(d_i, u_i) \right) & \text{if } w \circlearrowleft \text{ and (A.1)} \\ S_{\hat{\mathcal{C}}}(T_{\hat{\mathcal{C}}}(e, w, \overline{w}), \overline{w}) & \text{if } w \,_d\!\circlearrowleft \overline{w}. \end{cases}$$

where the definition of $S_{\hat{\mathcal{C}}}(\cdot, \cdot)$ makes use of the initialized relativized extraction function $T_{\hat{\mathcal{C}}}(\cdot, \cdot, \cdot)$, and induction on the well-founded relation $\leftarrow^+_{bo}$, noting that the first case in the definition of $S_{\hat{\mathcal{C}}}(\cdot, \cdot)$ assumes, like in Def. A.3, that the set of transitions of $\hat{\mathcal{C}}$ from $w$ is of the form (A.1) for some for $n_1, n_2, n_4 \in \mathbb{N}$, $a_1, \ldots, a_{n_1}, b_1, \ldots, b_{n_2}, d_1, \ldots, d_{n_4} \in A$, and $l_1, \ldots, l_{n_1}, l'_1, \ldots, l'_{n_2} \in \mathbb{N}$.

**Lemma A.5.** Let $\hat{\mathcal{C}}$ be a LLEE-witness of a chart $\mathcal{C} = \langle V, A, v_s, \to, \downarrow \rangle$. The following statements hold concerning preservation of under-star-1-free star expressions from arguments to values of the initialized relative extraction function $T_{\hat{\mathcal{C}}}$ of $\hat{\mathcal{C}}$, and the initialized extraction function $S_{\hat{\mathcal{C}}}$ of $\hat{\mathcal{C}}$:

(i) 1-free star expressions, and under-star-1-free star expressions, as initializing values for the initialized relative extraction function $T_{\hat{\mathcal{C}}}$ guarantee that the values of $T_{\hat{\mathcal{C}}}$ are again 1-free star expressions, and respectively, under-star-1-free star expressions. More formally, it holds:

$$\forall f \in StExp^{(1)}(A) \, \forall w, v \in V \bigl[ w \curvearrowleft v \implies T_{\hat{\mathcal{C}}}(f, w, v) \in StExp^{(1)}(A) \bigr], \tag{A.4}$$

$$\forall uf \in StExp^{(*/1)}(A) \, \forall w, v \in V \bigl[ w \curvearrowleft v \implies T_{\hat{\mathcal{C}}}(uf, w, v) \in StExp^{(*/1)}(A) \bigr]. \tag{A.5}$$

Equivalently, the restrictions of $T_{\hat{\mathcal{C}}}$ to the set $StExp^{(1)}(A)$ of 1-free star expressions and to the set $StExp^{(*/1)}(A)$ of under-star-1-free star expressions have the following types:

$$T_{\hat{\mathcal{C}}}(\cdot, \cdot, \cdot)|_{StExp^{(1)}(A)} : StExp^{(1)}(A) \times ((V \times V) \cap \curvearrowleft) \longrightarrow StExp^{(1)}(A)$$
$$\langle f, w, v \rangle \longmapsto T_{\hat{\mathcal{C}}}(f, w, v),$$

$$T_{\hat{\mathcal{C}}}(\cdot, \cdot, \cdot)|_{StExp^{(*/1)}(A)} : StExp^{(*/1)}(A) \times ((V \times V) \cap \curvearrowleft) \longrightarrow StExp^{(*/1)}(A)$$
$$\langle uf, w, v \rangle \longmapsto T_{\hat{\mathcal{C}}}(uf, w, v).$$

(ii) 1-Free star expressions, and under-star-1-free star expressions as initializing values for the initialized extraction function $S_{\hat{\mathcal{C}}}$ guarantee that the values of $S_{\hat{\mathcal{C}}}$ are again 1-free star expressions, and under-star-1-free star expressions, respectively. More formally, it holds:

$$\forall uf \in StExp^{(1)}(A) \, \forall w \in V \bigl[ S_{\hat{\mathcal{C}}}(uf, w) \in StExp^{(1)}(A). \bigr], \tag{A.6}$$

$$\forall uf \in StExp^{(*/1)}(A) \, \forall w \in V \bigl[ S_{\hat{\mathcal{C}}}(uf, w) \in StExp^{(*/1)}(A). \bigr]. \tag{A.7}$$

Equivalently, the restrictions of $S_{\hat{\mathcal{C}}}$ to the set $StExp^{(1)}(A)$ of under-star-1-free star expressions and to the set $StExp^{(*/1)}(A)$ have type:

$$S_{\hat{\mathcal{C}}}(\cdot, \cdot)|_{StExp^{(1)}(A)} : StExp^{(*/1)}(A) \times V \longrightarrow StExp^{(1)}(A)$$
$$\langle f, w \rangle \longmapsto S_{\hat{\mathcal{C}}}(f, w),$$

$$S_{\hat{\mathcal{C}}}(\cdot, \cdot)|_{StExp^{(*/1)}(A)} : StExp^{(*/1)}(A) \times V \longrightarrow StExp^{(*/1)}(A)$$
$$\langle uf, w \rangle \longmapsto S_{\hat{\mathcal{C}}}(uf, w).$$

*Proof.* Statements (i) and (ii) of the lemma can be proved by straightforward induction on the definitions of the initialized relative extraction function $S_{\hat{\mathcal{C}}}$ of $\hat{\mathcal{C}}$ in Def. A.3, and of the initialized extraction function $S_{\hat{\mathcal{C}}}$ of $\hat{\mathcal{C}}$ in Def. A.4, respectively.

By such proofs that proceed by the same kind of induction as used in Def. A.3, and respectively in Def. A.4, it can be observed: concerning (i), that, given 1-free star expressions $f$ as initializing values, values $T_{\hat{\mathcal{C}}}(f, w, v)$ of the initialized relative extraction function $T_{\hat{\mathcal{C}}}$ adhere to the grammar for 1-free star expressions in Def. 2.1; given under-star-1-free star expressions $uf$ as initializing values, values $T_{\hat{\mathcal{C}}}(uf, w, v)$ of the initialized relative extraction function $T_{\hat{\mathcal{C}}}$ adhere to the grammar for under-star-1-free star expressions in Def. 2.1; and concerning (ii), that, given 1-free star expressions $f$ (under-star-1-free star expressions $uf$) as initializing values, values $T_{\hat{\mathcal{C}}}(f, w, v)$ (resp., values $T_{\hat{\mathcal{C}}}(uf, w, v)$) of the initialized extraction function $T_{\hat{\mathcal{C}}}$ adhere to the grammar for 1-free star expressions (resp., under-star-1-free star expressions) in Def. 2.1. □

**Lemma A.6.** *Let $\mathcal{C} = \langle V, A, v_s, \rightarrow, \downarrow \rangle$ be a chart with LLEE-witness $\hat{\mathcal{C}}$. Then all values of the extraction function $s_{\hat{\mathcal{C}}}$ of $\hat{\mathcal{C}}$ are under-star-1-free star expressions, and consequently it has type: $s_{\hat{\mathcal{C}}} : V \rightarrow StExp^{(*/\mathbf{1})}(A)$.*

*Proof.* Due to $s_{\hat{\mathcal{C}}}(w) := S_{\hat{\mathcal{C}}}(1, w)$, for all $w \in V$, the fact that $s_{\hat{\mathcal{C}}}(w) \in StExp^{(*/\mathbf{1})}(A)$ follows from $1 \in StExp^{(*/\mathbf{1})}(A)$ and (A.7) in Lem. A.5, (ii). □

**Lemma A.7.** *Let $\mathcal{C} = \langle V, A, v_s, \rightarrow, \downarrow \rangle$ be a chart with LLEE-witness $\hat{\mathcal{C}}$. Then the initialized extraction function $S_{\hat{\mathcal{C}}}$ of $\hat{\mathcal{C}}$ has the following properties, for all $e \in StExp(A)$, and $v, w, w_1, w_2, w_3 \in V$:*

$$w \looparrowright v \implies T_{\hat{\mathcal{C}}}(e, w, v) \text{ is normed}, \tag{A.8}$$

$$v \rightsquigarrow w \not\looparrowright v \implies T_{\hat{\mathcal{C}}}(e, w, v) \text{ is not normed}, \tag{A.9}$$

$$w_3 \rightsquigarrow w_1 \looparrowright w_2 \looparrowright w_3 \implies T_{\hat{\mathcal{C}}}(e, w_1, w_3) = T_{\hat{\mathcal{C}}}(T_{\hat{\mathcal{C}}}(e, w_1, w_2), w_2, w_3), \tag{A.10}$$

$$w_2 \rightsquigarrow w_1 \not\looparrowright w_2 \land w_3 \rightsquigarrow w_1 \not\looparrowright w_3 \implies T_{\hat{\mathcal{C}}}(e, w_1, w_2) = T_{\hat{\mathcal{C}}}(e, w_1, w_3). \tag{A.11}$$

**Lemma A.8.** *Let $\mathcal{C} = \langle V, A, v_s, \rightarrow, \downarrow \rangle$ be a chart with LLEE-witness $\hat{\mathcal{C}}$. The initialized relative extraction function $T_{\hat{\mathcal{C}}}$, when initialized with an action $a \in A$, and the initialized extraction function $S_{\hat{\mathcal{C}}}$ of $\hat{\mathcal{C}}$, when initialized with an action $a \in A$ or with the values of the relative extraction function initialized with an action $a \in A$, enable the following $a$-transitions, for all $w, \overline{w} \in V$ with $\overline{w} \rightsquigarrow w$:*

$$T_{\hat{\mathcal{C}}}(a, w, \overline{w}) \xrightarrow{a} T_{\hat{\mathcal{C}}}(1, w, \overline{w}), \tag{A.12}$$

$$S_{\hat{\mathcal{C}}}(a, w) \xrightarrow{a} S_{\hat{\mathcal{C}}}(1, w), \tag{A.13}$$

$$S_{\hat{\mathcal{C}}}(T_{\hat{\mathcal{C}}}(a, w, \overline{w}), \overline{w}) \xrightarrow{a} \begin{cases} S_{\hat{\mathcal{C}}}(T_{\hat{\mathcal{C}}}(1, w, \overline{w}), \overline{w}) & \text{if } w \looparrowright \overline{w}, \\ T_{\hat{\mathcal{C}}}(1, w, \overline{w}) & \text{if } w \not\looparrowright \overline{w}. \end{cases} \tag{A.14}$$

*All of the transitions above are of the form $f \xrightarrow{a} uf$ for some $f \in StExp^{(\mathbf{1})}(A)$ and $uf \in StExp^{(*/\mathbf{1})}(A)$, and thus from 1-free star expressions to their derivatives that are under-star-1-free star expressions.*

**Lemma A.9.** *Let $\mathcal{C} = \langle V, A, v_s, \rightarrow, \downarrow \rangle$ be a chart with LLEE-witness $\hat{\mathcal{C}}$. Then the initialized relativized extraction function $T_{\hat{\mathcal{C}}}$ of $\hat{\mathcal{C}}$ has the following properties, for all $e \in StExp(A)$, and $w_1, w_2 \in V$:*

$$w_1 \looparrowright w_2 \implies S_{\hat{\mathcal{C}}}(e, w_1) = S_{\hat{\mathcal{C}}}(T_{\hat{\mathcal{C}}}(e, w_1, w_2), w_2), \text{ where } T_{\hat{\mathcal{C}}}(e, w_1, w_2) \text{ is normed}. \tag{A.15}$$

$$w_2 \rightsquigarrow w_1 \not\looparrowright w_2 \implies S_{\hat{\mathcal{C}}}(e, w_1) = T_{\hat{\mathcal{C}}}(e, w_1, w_2), \text{ which is not normed}. \tag{A.16}$$

**Lemma A.10** (corresponds to statement $(\mathbf{E})_{\mathbf{1}}^{(\mathrm{ref})}$). *Let $\mathcal{C} = \langle V, A, v_s, \rightarrow, \downarrow \rangle$ be a chart with LLEE-witness $\hat{\mathcal{C}}$. Then the solution $s_{\hat{\mathcal{C}}} : V \rightarrow StExp^{(*/\mathbf{1})}(A)$ of $\mathcal{C}$ that is extracted from $\hat{\mathcal{C}}$ defines a (functional) bisimulation from $\mathcal{C}$ to the chart interpretation $\mathcal{C}(s_{\hat{\mathcal{C}}}(v_s))$ of $s_{\hat{\mathcal{C}}}(v_s)$, and hence $\mathcal{C} \rightleftharpoons_{s_{\hat{\mathcal{C}}}} \mathcal{C}(s_{\hat{\mathcal{C}}}(v_s))$.*

*Proof (hint).* The statement of the lemma can be established by arguing, for all $w \in V$, along the case distinction in the definition of $s_{\hat{\mathcal{C}}}(w) = S_{\hat{\mathcal{C}}}(1, w)$ to show that, for each $a \in A$, the $a$-derivatives of $s_{\hat{\mathcal{C}}}(w)$ coincide with the values of $s_{\hat{\mathcal{C}}}$ that are reached via, in the definition of $s_{\hat{\mathcal{C}}}$ corresponding, $a$-transitions from $w$. In the argument, Lem. A.7, Lem. A.8, and Lem. A.9 are used. □

**Proposition A.11.** *Let $\mathcal{C} = \langle V, A, v_s, \rightarrow, \downarrow \rangle$ be a chart with LLEE-witness $\hat{\mathcal{C}}$ such that $\mathcal{C}$ is a bisimulation collapse. Then the solution $s_{\hat{\mathcal{C}}} : V \rightarrow StExp^{(*/\mathbf{1})}(A)$ of $\mathcal{C}$ that is extracted from $\hat{\mathcal{C}}$ defines an isomorphism from $\mathcal{C}$ to the chart interpretation $\mathcal{C}(s_{\hat{\mathcal{C}}}(v_s))$ of $s_{\hat{\mathcal{C}}}(v_s)$, and hence $\mathcal{C} \simeq \mathcal{C}(s_{\hat{\mathcal{C}}}(v_s))$.*

*Proof.* By Lemma A.10, $s_{\hat{\mathcal{C}}}$ defines a functional bisimulation from $\mathcal{C}$ to $\mathcal{C}(s_{\hat{\mathcal{C}}}(v_s))$. If $\mathcal{C}$ is collapsed, it follows that also $\mathcal{C}(s_{\hat{\mathcal{C}}}(v_s))$ must be collapsed, and that $s_{\hat{\mathcal{C}}}$ defines an isomorphism between $\mathcal{C}$ and $\mathcal{C}(s_{\hat{\mathcal{C}}}(v_s))$. □

*Proof of Prop. 2.12.* Suppose that a chart $\mathcal{C}$ is in the image of the chart interpretation of the set of under-star-1-free star expressions over $A$. That means, $\mathcal{C} = \mathcal{C}(e)$ for an under-star-1-free star expression $e \in StExp^{(*/\mathbf{1})}(A)$. We pick $e$ accordingly. We let $\mathcal{C}_0 = \langle V_0, A, v_{s,0}, \rightarrow, \downarrow \rangle$ be a bisimulation collapse of $\mathcal{C}$. We have to show that $\mathcal{C}_0$ is in the image of the chart interpretation of the set of under-star-1-free star expressions.

From $\mathcal{C} = \mathcal{C}(e)$ it follows by $(\mathbf{I})_{\mathbf{1}}$ that $\mathcal{C}$ is a LLEE-chart. By $(\mathbf{C})$ we get that also the bisimulation collapse $\mathcal{C}_0$ of $\mathcal{C}$ is a LLEE-chart. Therefore we can pick a LLEE-witness $\hat{\mathcal{C}}_0$ of $\mathcal{C}_0$. Then due to $(\mathbf{E})_{\mathbf{1}}^{(\mathrm{ref})}$, as also stated by Prop. A.11 it follows that $\mathcal{C}_0 \simeq \mathcal{C}(s_{\hat{\mathcal{C}}_0}(v_{s,0}))$, and $uf_0 := s_{\hat{\mathcal{C}}_0}(v_{s,0}) \in StExp^{(*/\mathbf{1})}(A)$ is an under-star-1-free star expression. Therefore $\mathcal{C}_0 \simeq \mathcal{C}(uf_0) \in \mathcal{C}(StExp^{(*/\mathbf{1})}(A))$ holds, and thus the bisimulation collapse $\mathcal{C}_0$ of $\mathcal{C}$ is in the image $\mathcal{C}(StExp^{(*/\mathbf{1})}(A))$ of the chart interpretation $\mathcal{C}(\cdot)$ of the set $StExp^{(*/\mathbf{1})}(A)$ of under-star-1-free star expressions over $A$, modulo isomorphism. □

### A.2. Supplements for Section 3: Refined process semantics

**Lemma A.12** (= Lem. 3.8). *Derivability of statements concerning termination, and transitions in $\mathcal{T}(A)$, and in $\underline{\mathcal{T}}(A)$ are related as follows, for all $e, e' \in StExp(A)$, and $a \in A$:*

$$\vdash_{\mathcal{T}(A)} e\!\downarrow \iff \vdash_{\underline{\mathcal{T}}(A)} e \xrightarrow{1}{}^* 1\,, \tag{A.17}$$

$$\vdash_{\mathcal{T}(A)} e \xrightarrow{a} e' \iff \vdash_{\underline{\mathcal{T}}(A)} e \xrightarrow{1}{}^* \cdot \xrightarrow{a} e'\,. \tag{A.18}$$

For the proof of this lemma we use an extension of the TSS $\underline{\mathcal{T}}(A)$ by rules that permit to describe induced transitions, and induced termination with respect to $\mathcal{T}(A)$.

**Definition A.13.** We denote by $\underline{\mathcal{T}}_{(\cdot]}(A)$ *the TSS for induced termination, and for induced transitions with respect to $\underline{\mathcal{T}}(A)$* that results by adding, to the axioms and rules of $\underline{\mathcal{T}}(A)$, the following four rules:

$$\frac{}{1\!\downarrow^{(1)}} \qquad \frac{e \xrightarrow{1} \tilde{e} \quad \tilde{e}\!\downarrow^{(1)}}{e\!\downarrow^{(1)}} \qquad \frac{e \xrightarrow{a} e'}{e \xrightarrow{(a]} e'} \qquad \frac{e \xrightarrow{1} \tilde{e} \quad \tilde{e} \xrightarrow{(a]} e'}{e \xrightarrow{(a]} e'}$$

With respect to this extension of $\underline{\mathcal{T}}(A)$ we can reformulate, and prove more easily, the statement of Lem. 3.8 as follows.

**Lemma A.14.** *Derivability of statements concerning termination, and transitions in $\mathcal{T}(A)$, and in $\underline{\mathcal{T}}_{(\cdot]}(A)$ are related as follows, for all $e, e' \in StExp(A)$, and $a \in A$:*

$$\vdash_{\mathcal{T}(A)} e\!\downarrow \iff \vdash_{\underline{\mathcal{T}}_{(\cdot]}} e\!\downarrow^{(1)}\,, \tag{A.19}$$

$$\vdash_{\mathcal{T}(A)} e \xrightarrow{a} e' \iff \vdash_{\underline{\mathcal{T}}_{(\cdot]}(A)} e \xrightarrow{(a]} e'\,. \tag{A.20}$$

*Proof (method).* The directions "⇒" in (A.19) and in (A.20) can be established with easy proofs by induction on the depth of derivations with conclusion $e\!\downarrow$ and with conclusion $e \xrightarrow{a} e'$ in $\mathcal{T}(A)$, thereby transforming these derivations step by step into derivations with conclusion $e\!\downarrow^{(1)}$ and with conclusion $e \xrightarrow{(a]} e'$ in $\underline{\mathcal{T}}_{(\cdot]}(A)$, respectively.

Vice versa, the directions "⇐" in (A.19) and in (A.20) can be established with proofs by induction on the depth of derivations with conclusion $e\!\downarrow^{(1)}$ and with conclusion $e \xrightarrow{(a]} e'$ in $\underline{\mathcal{T}}_{(\cdot]}(A)$, thereby transforming these derivations step by step into derivations with conclusion $e\!\downarrow$ and with conclusion $e \xrightarrow{a} e'$ in $\mathcal{T}(A)$, respectively. □

On the basis of this additional lemma, Lem. A.12 can be established easily indeed.

*Proof (of Lem. A.12).* Derivability of $e\!\downarrow^{(1)}$ in $\underline{\mathcal{T}}_{(\cdot]}(A)$ is equivalent to derivability of $e \xrightarrow{1}{}^* 1$ in $\underline{\mathcal{T}}(A)$, for all $e \in StExp(A)$. Also, derivability of $e \xrightarrow{(a]} e'$ in $\underline{\mathcal{T}}_{(\cdot]}(A)$ is equivalent to derivability of $e \xrightarrow{1}{}^* \cdot \xrightarrow{a} e'$ in $\underline{\mathcal{T}}(A)$. These two statements can be proved by straightforward proofs by induction on the depth of derivations.

By applying these two statements that link derivability in $\underline{\mathcal{T}}_{(\cdot]}(A)$ with derivability in $\mathcal{T}(A)$, the statements (A.17) and (A.18) of Lem. A.12 follow from the statements (A.19) and (A.20) of Lem. A.14, respectively. □

**Lemma** (= Lem. 3.9). *For every $e \in StExp(A)$, the 1-chart interpretation $\underline{\mathcal{C}}(e)$ of $e$ is a finite, weakly guarded 1-chart.*

*Proof.* The size $sz(e)$ of a star expression $e$ (that is, the number of its symbols) decreases strictly along 1-transitions, that is, if $e_1 \xrightarrow{1} e_2$ is a 1-transition that is specified by the TSS $\underline{\mathcal{T}}(A)$, then $sz(e_1) > sz(e_2)$ holds. This can be verified by a straightforward induction on derivations in $\underline{\mathcal{T}}(A)$ with 1-transitions in their conclusions. Consequently, the 1-chart interpretation $\underline{\mathcal{C}}(e)$ of a star expression $e$ cannot contain an infinite path of 1-transitions. Therefore $\underline{\mathcal{C}}(e)$ is weakly guarded, for every star expression $e$.

$\underline{\mathcal{C}}(e)$ is also finitely branching, by a similar argument. The actions that occur in derivatives of a star expression $e_1$ are among those that occur in $e_1$. Since the size of an expression decreases properly in every 1-transition $e_1 \xrightarrow{1} e_2$, it follows that targets $e_2$ of such 1-transitions are among the finitely many star expressions of size smaller than $sz(e_1)$ in which only actions occur that also occur in $e_1$. Therefore there can be only finitely many 1-transitions from a star expression $e_1 \in \underline{\mathcal{T}}(A)$.

Finiteness of the 1-chart interpretation $\underline{\mathcal{C}}(e)$ follows from weak guardedness, for all $e \in StExp(A)$, due to the following two facts: first, $\underline{\mathcal{C}}(e)$ 1-transition refines the chart interpretation $\mathcal{C}(e)$ of $e$, because induced transitions with respect to $\underline{\mathcal{T}}(A)$ refine transitions with respect to $\mathcal{T}(A)$ due to (A.18); and second, the chart interpretation $\mathcal{C}(e)$ of a star expression $e$ is finite due to Lem. 2.5. It follows that the vertex set of $\underline{\mathcal{C}}(e)$ consists of the vertices of $\mathcal{C}(e)$ plus all vertices that are reachable from these vertices by 1-transitions. Since due to weak guardedness there can be no infinite 1-transition paths, and the 1-transition relation is finitely branching, it follows by König's Lemma that for every vertex $v$ of $\mathcal{C}(e)$ there can be only finitely many more vertices that are added in $\underline{\mathcal{C}}(e)$ by 1-transitions to additional vertices that are not proper-transition targets. (Note that proper-transition targets of $\underline{\mathcal{C}}(e)$ are also vertices of $\mathcal{C}(e)$ due to the refinement property (A.18).) □

**Lemma** (= Lem. 3.15). *The 1-transition elimination rewrite relation $\to_{(1)}$ has the following properties, for all 1-charts $\underline{\mathcal{C}}, \underline{\mathcal{C}}_1, \underline{\mathcal{C}}_2$:*
  *(i) If $\underline{\mathcal{C}}_1$ is weakly guarded, and $\underline{\mathcal{C}}_1 \to^*_{(1)} \underline{\mathcal{C}}_2$, then $\mathcal{C}_2$ is finite and w.g., and $|V(\mathcal{C}_1)| - 1 \leq |V(\mathcal{C}_2)| \leq |V(\mathcal{C}_1)|$.*
  *(ii) $\to_{(1)}$ is terminating from every finite w.g. 1-chart.*
  *(iii) $\to_{(1)}$ normal forms are 1-free 1-chart. $\to_{(1)}$ normal forms of finite, w.g. 1-charts are finite 1-free 1-charts.*
  *(iv) $\underline{\mathcal{C}}_1 \to^*_{(1)} \underline{\mathcal{C}}_2 \implies (\underline{\mathcal{C}}_1)_{(\cdot]} = (\underline{\mathcal{C}}_2)_{(\cdot]}$.*
  *(v) If $\underline{\mathcal{C}}$ is finite and weakly guarded, and $\mathcal{C}$ is 1-free, then:*
     $\underline{\mathcal{C}} \to^*_{(1)} \mathcal{C} \iff (\underline{\mathcal{C}})_{(\cdot]} = \mathcal{C}$.
  *(vi) If $\underline{\mathcal{C}}$ is finite, and weakly guarded, then $\underline{\mathcal{C}} \to^*_{(1)} \underline{\mathcal{C}}_{(\cdot]}$, that is, $\underline{\mathcal{C}}$ refines its induced chart $\underline{\mathcal{C}}_{(\cdot]}$, and $\underline{\mathcal{C}}_{(\cdot]}$ is the unique $\to_{(1)}$ normal form of $\underline{\mathcal{C}}$.*

*Proof.* For (i) it suffices to note: every $\to_{(1)}$ step can only shorten 1-transition paths, but it cannot introduce a 1-transition cycle; in every $\to_{(1)}$ step at most one vertex, the target $v$ of the 1-transition that is removed, can become unreachable.

For (ii) it suffices to obtain a measure $m(\underline{\mathcal{C}})$ on finite w.g. 1-charts $\underline{\mathcal{C}}$ that decreases properly in every $\to_{(1)}$ step from $\underline{\mathcal{C}}$. We can use $m(\underline{\mathcal{C}}) := \sum_{v \in V(\underline{\mathcal{C}})} s_{\underline{\mathcal{C}}}(v)$ where $s_{\underline{\mathcal{C}}}(v)$ denotes the sum of the lengths of all maximal 1-transition paths from $v$. Note that, for both rules in Def. 3.13, the length of every maximal 1-transition path from $v_0$ decreases by 1, whereas maximal 1-transition paths from other vertices are either preserved, or are also shortened by 1 if they pass through $v_0$.

For (iii) we note that every 1-transition gives rise to a $\to_{(1)}$ step, so normal forms do not contain 1-transitions, and that by (i), (ii) every finite w.g. 1-chart rewrites via finitely many $\to_{(1)}$ steps to a normal form that is finite and w.g..

Statement (iv) can be shown by induction on the length of $\to_{(1)}$ paths by using that every $\to_{(1)}$ step preserves induced transitions and induced termination. The direction "$\Rightarrow$" in (v) follows from (iv) and that $\mathcal{C}_{(\cdot]} = \mathcal{C}$ for 1-free 1-charts $\mathcal{C}$. For showing the direction "$\Leftarrow$" in (v), suppose that $(\underline{\mathcal{C}})_{(\cdot]} = \mathcal{C}$ for finite w.g. 1-charts $\underline{\mathcal{C}}, \mathcal{C}$ where $\mathcal{C}$ is 1-free. We have to show $\underline{\mathcal{C}}'_{(\cdot]} = \mathcal{C}$. Then $\underline{\mathcal{C}} \to^*_{(1)} \underline{\mathcal{C}}'$ for an $\to_{(1)}$ normal form $\underline{\mathcal{C}}'$ by (ii), which is finite, w.g., and 1-free by (iii). Then we get $\underline{\mathcal{C}}'_{(\cdot]} = \underline{\mathcal{C}}_{(\cdot]} = \mathcal{C}$ by using (iv), and thus obtain $\underline{\mathcal{C}}'_{(\cdot]} = \mathcal{C}$.

Last, (vi) follows from "$\Leftarrow$" in (v), letting $\mathcal{C} := \underline{\mathcal{C}}_{(\cdot]}$, and by recalling (iii). □

### A.3. Supplements for Section 4: LLEE-witnesses for 1-charts

**Theorem** (= Thm. 4.7). *The entry/body-labeling $\widehat{\underline{\mathcal{C}}(e)}$ of $\underline{\mathcal{C}}(e)$ is a LLEE-witness of the 1-chart interpretation $\underline{\mathcal{C}}(e)$ of $e$, for every $e \in StExp(A)$. Therefore the 1-chart interpretation $\underline{\mathcal{C}}(e)$ of a star expression $e \in StExp(A)$ satisfies LEE.*

The proof of this theorem can be given as an adaptation of the proof of a similar statement that is proved in [13], [19]. There, a different version of the 1-chart interpretation of star expressions is defined together with an entry/body-labeling (see Rem. 4.11) that is proved to be a LLEE-witness. We only sketch the most important steps and lemmas of the adaptation.

Thm. 4.7 states that the entry/body-labeling $\widehat{\underline{\mathcal{C}}(e)}$ from Def. 4.5 is a LLEE-witness for the 1-chart interpretation $\underline{\mathcal{C}}(e)$ of $e$. The proof of the theorem can be assembled from three auxiliary statements, Lemma A.15, Lemma A.16, and Lemma A.17 below. For the formulation of these lemmas, we introduce the set *AppCxt*($A$) of *applicative contexts of stacked star expressions* over $A$ by which we mean the set of contexts that are defined by the grammar:

$$C[\cdot] ::= \Box \mid C[\cdot] \cdot e \quad \text{(where } e \in StExp(A)\text{)}.$$

The first lemma states that transitions with normed targets of entry/body-labelings defined by the TSS $\widehat{\underline{\mathcal{T}}}(A)$ in Def. 4.5 are preserved under the operation of filling star expressions into applicative contexts. Lem. A.16 describes body transition paths from product expressions filled in applicative contexts, and Lem. A.17 gathers important properties of paths and steps of transitions.

**Lemma A.15.** *If $e'$ is normed, and $e \to_l e'$ is derivable in $\widehat{\underline{\mathcal{T}}}(A)$, then so is $C[e] \to_l C[e']$, for every $l \in \{\text{bo}\} \cup \{[n] \mid n \in \mathbb{N}^+\}$.*

*Proof.* By induction on the structure of applicative contexts, using the rules for $\cdot$ in $\widehat{\underline{\mathcal{T}}}(A)$. □

**Lemma A.16.** *Every maximal $\to_{bo}$ path according to $\widehat{\underline{\mathcal{T}}}(A)$, for $e, f \in StExp(A)$, is of either of the following three forms:*

(i) $C[e \cdot f] = C[e_0 \cdot f] \to_{bo} C[e_1 \cdot f] \to_{bo} \ldots \to_{bo} C[e_n \cdot f] \to_{bo} \ldots$ *is finite or infinite*[2] *with normed star expressions $e_0, e_1, \ldots, e_n, \ldots \in StExp(A)$ such that $e_0 \to_{bo} e_1 \to_{bo} \ldots \to_{bo} e_n \to_{bo} \ldots$,*

(ii) $C[e \cdot f] = C[e_0 \cdot f] \to_{bo} C[e_1 \cdot f] \to_{bo} \ldots \to_{bo} C[e_n \cdot f] = C[1 \cdot f] \to_{bo} C[f] \to_{bo} \ldots$ *with $n \in \mathbb{N}$, and normed star expressions $e_0, e_1, \ldots, e_n \in StExp(A)$ such that $e_0 \to_{bo} e_1 \to_{bo} \ldots \to_{bo} e_n = 1$,*

(iii) $C[e \cdot f] = C[e_0 \cdot f] \to_{bo} C[e_1 \cdot f] \to_{bo} \ldots \to_{bo} C[e_n \cdot f] \to_{bo} e_{n+1} \to_{bo} \ldots$ *with $n \in \mathbb{N}$, and normed star expressions $e_0, e_1, \ldots, e_n \in StExp(A)$ and not normed $e_{n+1}$ such that $e_0 \to_{bo} e_1 \to_{bo} \ldots \to_{bo} e_n \to_{bo} e_{n+1} \to_{bo} \ldots$.*

**Lemma A.17.** *The following statements hold for paths of transitions that are induced by the TSS $\widehat{\underline{\mathcal{T}}}(A)$:*

(i) *There are no infinite $\to_{bo}$ paths.*

(ii) *If $e \in StExp(A)$ is normed, then $e \to_{bo}^* 1$.*

(iii) *If $e \to_{[n]} e'$ with $n > 0$ and $e, e' \in StExp(A)$, then $e = C[f^*]$, $f \to_l f_0'$, $e' = C[f_0' \cdot f^*]$, and $n = |f|_* + 1$, for some $f \in StExp(A)$, $C[\cdot] \in AppCxt(A)$, and normed $f_0' \in StExp(A)$.*

(iv) *Neither $\to_{bo}$ and $\to_{[n]}$ steps, where $n \geq 1$, increase the (syntactic) star height of expressions.*

*Proof of Thm. 4.7 (outline).* The goal is to show that the entry/body-labeling $\widehat{\mathcal{C}(g)}$ defined by the TSS $\widehat{\underline{\mathcal{T}}}(A)$ in Def. 4.5 is a LLEE-witness of the 1-chart interpretation $\underline{\mathcal{C}}(g)$, for every star expression $g \in StExp(A)$.

Instead of verifying the LLEE-witness conditions (W1), (W2), (W3) for all $\widehat{\mathcal{C}(g)}$ where $g \in StExp(A)$ is arbitrary, the following four general statements can be shown, from which the proof goal follows (see below). The conditions below are understood to be universally quantified over all $e, e_1, f, f_1 \in StExp(A)$, and $n, m \in \mathbb{N}$ with $n, m > 0$:

(LLEE-1) $e \to_{[n]} e_1 \implies e_1 \to_{bo}^* e$,

(LLEE-2) $\to_{bo}$ is terminating from $e$,

(LLEE-3) $e \xrightarrow{\phantom{a}}_{\not\in(e)}{}_{[n]} \cdot \xrightarrow{\phantom{a}}_{\not\in(e)}{}_{bo}^* f \implies f \neq 1$ (the premise means that $f$ is in $\underline{\mathcal{C}}_{\widehat{\mathcal{C}(g)}}(e, n)$ such that $f \neq e$),

(LLEE-4) $e \xrightarrow{\phantom{a}}_{\not\in(e)}{}_{[n]} \cdot \xrightarrow{\phantom{a}}_{\not\in(e)}{}_{bo}^* f \to_{[m]} f_1 \implies n > m$,

Hereby $\xrightarrow{\phantom{a}}_{\not\in(e)}{}_{[n]}$ and $\xrightarrow{\phantom{a}}_{\not\in(e)}{}_{bo}$ means $\to_{[n]}$ and $\to_{bo}$ steps, respectively, that avoid $e$ as their targets.

These four conditions imply that $\widehat{\mathcal{C}(g)}$ is a LLEE-witness, for all $g \in StExp(A)$. Let $g \in StExp(A)$ be arbitrary. Then (LLEE-2) obviously implies (W1) for $\widehat{\mathcal{C}(g)}$. For each entry identifier $\langle e, n \rangle \in E(\widehat{\mathcal{C}(g)})$ it is straightforward to check that the statements (LLEE-1), (LLEE-2), and (LLEE-3) imply that $\underline{\mathcal{C}}_{\widehat{\mathcal{C}(g)}}(e, n)$ satisfies the loop properties (L1), (L2), and (L3), respectively, to obtain (W2) for $\widehat{\mathcal{C}(g)}$ Finally, (LLEE-4) clearly implies the condition (W3) for $\widehat{\mathcal{C}(g)}$.

Then it is not difficult to verify the conditions (LLEE-1)–(LLEE-4) above. The arguments follow closely the proof of Lemma 5.13 in [13], [19], using the slightly different versions of the last two lemmas above, Lem. A.16 and Lem. A.17.

### A.4. Supplements for Section 5: Counterexample

**Lemma** (= Lem. 5.5). *For every finite 1-chart $\underline{\mathcal{D}}$ with LLEE–1-lim there is a finite 1-chart $\underline{\mathcal{D}}'$ with the same induced chart, that is, $\underline{\mathcal{D}}'_{(\cdot]} = \underline{\mathcal{D}}_{(\cdot]}$, and with a 1-transition limited LLEE-witness $\hat{\underline{\mathcal{D}}}'$ such that it holds:*

(ptt) *If for a loop-entry identifier $\langle v, n \rangle \in E(\hat{\underline{\mathcal{D}}}')$ the loop sub-1-chart $\underline{\mathcal{C}}_{\hat{\underline{\mathcal{D}}}'}(v, n)$ of $\underline{\mathcal{D}}'$ (a) consists of a single scc, (b) contains no other loop vertex than $v$ ($v$ is innermost), then it contains only proper-transition targets in $\underline{\mathcal{D}}'$.*

*Proof of Lem. 5.5.* It suffices to show that in a finite 1-chart $\underline{\mathcal{D}}'$ with LLEE–1-lim, and with a 1-transition limited LLEE-witness $\hat{\underline{\mathcal{D}}}'$ every violation of (ptt) for a loop-entry identifier $\langle v, n \rangle \in E(\hat{\underline{\mathcal{D}}}')$ can be removed with as result a finite 1-chart $\underline{\mathcal{D}}'$ with a 1-transition limited LLEE-witness $\hat{\underline{\mathcal{D}}}'$ that has one violation of (ptt) less than $\underline{\mathcal{D}}$ and $\hat{\underline{\mathcal{D}}}$, and that has the same induced chart as $\underline{\mathcal{D}}$, that is, $\underline{\mathcal{D}}'_{(\cdot]} = \underline{\mathcal{D}}_{(\cdot]}$. This is because repeated application of this transformation statement to a 1-chart with LLEE–1-lim $\underline{\mathcal{D}}$ then yields a LLEE-1-chart $\underline{\mathcal{D}}'$ and a LLEE-witness $\hat{\underline{\mathcal{D}}}'$ with (ptt) and with the same induced chart, as stated by the lemma.

Suppose that $\underline{\mathcal{D}}$ is a 1-chart with 1-transition limited LLEE-witness $\hat{\underline{\mathcal{D}}}$ such that there is a violation of (ptt) in the form of a loop-entry identifier $\langle v, n \rangle \in E(\hat{\underline{\mathcal{D}}}')$ such that $\underline{\mathcal{C}}_{\hat{\underline{\mathcal{D}}}}(v, n)$ (a) consists of a single scc, (b) is innermost, but (c) it contains vertices that are not proper-transition targets in $\underline{\mathcal{D}}$.

We first show that $v$ is the single vertex of $\underline{\mathcal{C}}_{\hat{\underline{\mathcal{D}}}}(v, n)$ that is not a proper-transition target in $\underline{\mathcal{D}}$. For this, let $u$ be a vertex of $\underline{\mathcal{C}}_{\hat{\underline{\mathcal{D}}}}(v, n)$ that is not a proper-transition target. Then $u$ must be the target of a 1-transition. But as 1-transitions are back-links in $\hat{\underline{\mathcal{D}}}$ since $\hat{\underline{\mathcal{D}}}$ is 1-transition limited, it follows that $u$ must be a loop vertex of $\hat{\underline{\mathcal{D}}}$. However, as $v$ is an innermost loop vertex of $\widehat{\underline{\mathcal{D}}}$ by assumption (b), we conclude that $u = v$. Since $\mathcal{C}_{\widehat{\underline{\mathcal{D}}}}(v, n)$ contains vertices that are not proper-transition targets by assumption (c), it follows that $v$ must be such a vertex. Consequently it is the single vertex in $\mathcal{C}_{\widehat{\underline{\mathcal{D}}}}(v, n)$ that is not a proper-transition target.

---

2. But note that Lem. A.17, (i), below then excludes infinite $\to_{bo}$ paths.

Next we note that $\mathcal{C}_{\widehat{\underline{\mathcal{D}}}}(v, n)$ cannot contain a 1-transition self-loop at $v$, because $\underline{\mathcal{D}}$ is weakly guarded as a 1-chart with LLEE–1-lim. Therefore all transitions from $v$ in $\mathcal{C}_{\widehat{\underline{\mathcal{D}}}}(v, n)$ are proper transitions. We choose a maximal path $\pi$ of proper transitions from $v$ in $\mathcal{C}_{\widehat{\underline{\mathcal{D}}}}(v, n)$. This path $\pi$ departs from $v$ in the first step, and does not return to $v$, because otherwise $v$ would be a proper-transition target. The path $\pi$ must be finite, due to loop 1-chart condition (L2) on the loop sub-1-chart $\mathcal{C}_{\widehat{\underline{\mathcal{D}}}}(v, n)$ of $\underline{\mathcal{D}}$. Therefore $\pi$ stops at a vertex $w$ from which no further proper transition departs. However, since $\mathcal{C}_{\widehat{\underline{\mathcal{D}}}}(v, n)$ is an scc by assumption (a), there must be an outgoing 1-transition from $w$. As $\underline{\mathcal{D}}$ is 1-transition limited, this 1-transition must be a back-link to $v$. In the example below left, for 1-chart $\underline{\mathcal{D}} = \underline{\mathcal{D}}_1$ with LLEE-witness $\widehat{\underline{\mathcal{D}}}_1$, this vertex $w$ is unique, but there are three possible paths from $v$ to $w$. In the example below right, for 1-chart $\underline{\mathcal{D}} = \underline{\mathcal{D}}_2$ with LLEE-witness $\widehat{\underline{\mathcal{D}}}_2$, there are two possible choices for $w$ (of which only one is drawn).

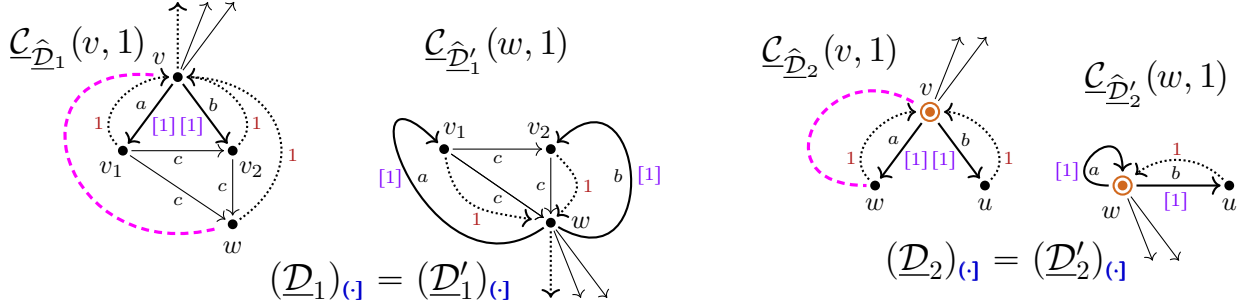

Since the only outgoing transition from $w$ in $\underline{\mathcal{D}}$ is a 1-transition to $v$ (by construction of $w$), the vertices $w$ and $v$ are 1-bisimilar in $\underline{\mathcal{D}}$ (as indicated by the magenta links in the pictures).

We can now transform $\underline{\mathcal{D}}$ and $\widehat{\underline{\mathcal{D}}}$ by removing $v$, and by letting the loop sub-1-chart start at $w$ instead, transferring the loop-entry transitions in $\widehat{\underline{\mathcal{D}}}$ from $v$ to $w$, directing 1-transition back-links in $\widehat{\underline{\mathcal{D}}}$ to $v$ now over to $w$, and moving possible termination at $v$ over to $w$ (see the second example above), as well as also changing the source of transitions that depart from $v$ to $w$. In this way we obtain from $\underline{\mathcal{D}}$ and $\widehat{\underline{\mathcal{D}}}$ a 1-chart $\underline{\mathcal{D}}'$ with a 1-transition limited LLEE-witness $\widehat{\underline{\mathcal{D}}}'$ such that:

▷ induced transitions in $\underline{\mathcal{D}}$ are preserved as induced transitions of $\widehat{\underline{\mathcal{D}}}$,
▷ therefore $\underline{\mathcal{D}}'_{(\cdot)} = \underline{\mathcal{D}}_{(\cdot)}$ follows,
▷ now $\mathcal{C}_{\widehat{\underline{\mathcal{D}}}'}(v, n)$ consists only of proper-transition targets,
▷ other innermost loop sub-1-charts of $\underline{\mathcal{D}}$ with respect to $\widehat{\underline{\mathcal{D}}}$ are not changed.

In this way we have shown the proof obligation for the lemma. □

**Lemma** (= Lem. 5.6). *Neither of the 1-charts $\underline{\mathcal{C}}_1$ and $\underline{\mathcal{C}}_{10}$ in Ex. 5.1 can be refined into a wg-LLEE-1-chart.*

*Part of the proof of Lem. 5.6 for $\underline{\mathcal{C}}_1$ that is analogous to the argument for $\underline{\mathcal{C}}_2$ in the submission.* In the proof of Lem. 5.6 we had to show, for the 1-charts $\underline{\mathcal{C}}_1$ in Ex. 5.3, and $\underline{\mathcal{C}}_2$ defined in the proof, in particular:

(S2) Neither $\underline{\mathcal{C}}_1$ nor $\underline{\mathcal{C}}_2$ can be refined into a 1-chart with LLEE–1-lim.

We argued that for $\underline{\mathcal{C}}_2$, and stated that the argument is analogous for $\underline{\mathcal{C}}_1$. Here we demonstrate that that is indeed the case.

We now assume that a 1-chart $\underline{\mathcal{D}}$ with 1-transition limited LLEE-witness $\widehat{\underline{\mathcal{D}}}$ is a 1-transition refinement of $\underline{\mathcal{C}}_1$. Then $\underline{\mathcal{D}} \rightarrow^+_{(1)} \underline{\mathcal{C}}_1$ follows, because $\underline{\mathcal{C}}_1$ is not a LLEE-1-chart (see Lem. 5.4) contrary to $\underline{\mathcal{D}}$.

Now we note that $\underline{\mathcal{C}}_1$ cannot be refined in the direction of the 1-transition limited LLEE-witness $\widehat{\underline{\mathcal{D}}}$ by adding 1-transition back-links to vertices that are already present in $\underline{\mathcal{C}}_1$. This is because, in order to satisfy the layeredness condition (W3) for the LLEE-witness $\widehat{\underline{\mathcal{D}}}$, any such back-link needed to start from $v$, or from a vertex outside of $\mathcal{C}_{\widehat{\underline{\mathcal{D}}}}(v, 1)$ (the sub-1-chart of $\underline{\mathcal{C}}_1$ delimited by $\widehat{\underline{\mathcal{D}}}$ and by its entry/body-labeling $\widehat{\underline{\mathcal{C}}}_1$), and target a different of these vertices that furthermore needs to be a substate. But there is no such a vertex in $\underline{\mathcal{C}}_2$: neither $v$ nor one of $\overline{w}_1$, $\boxed{c_1}$, $\boxed{c_2}$, $\boxed{f}$ outside of $\mathcal{C}_{\widehat{\underline{\mathcal{D}}}}(v, 1)$ is a substate of either of these five vertices in $\underline{\mathcal{C}}_1$.

Consequently, the only option to refine $\widehat{\underline{\mathcal{C}}}_1$ towards the 1-transition limited LLEE-witness $\widehat{\underline{\mathcal{D}}}$ is to add a vertex $u$ that is not a proper-transition target, with 1-transition back-links directed to it from one of the five vertices $v$, $\overline{w}_1$, $\boxed{c_1}$, $\boxed{c_2}$, or $\boxed{f}$, and such that proper transitions in $\underline{\mathcal{C}}_1$ arise as induced transitions via $u$. The only option to simplify the structure is to share, via the new vertex $u$ proper transitions from two of these five vertices. Such a non-trivial sharing of transitions is possible only between $\overline{w}_1$ and $v$, because none of other pairs of vertices have an action label of an outgoing transition in common. The maximal option is to share, via the new vertex $u$, all $a$- and $c$-transitions that depart from $v$ and from $\overline{w}_1$. This leads to the 1-chart $\underline{\mathcal{C}}'_1$ on the left below:

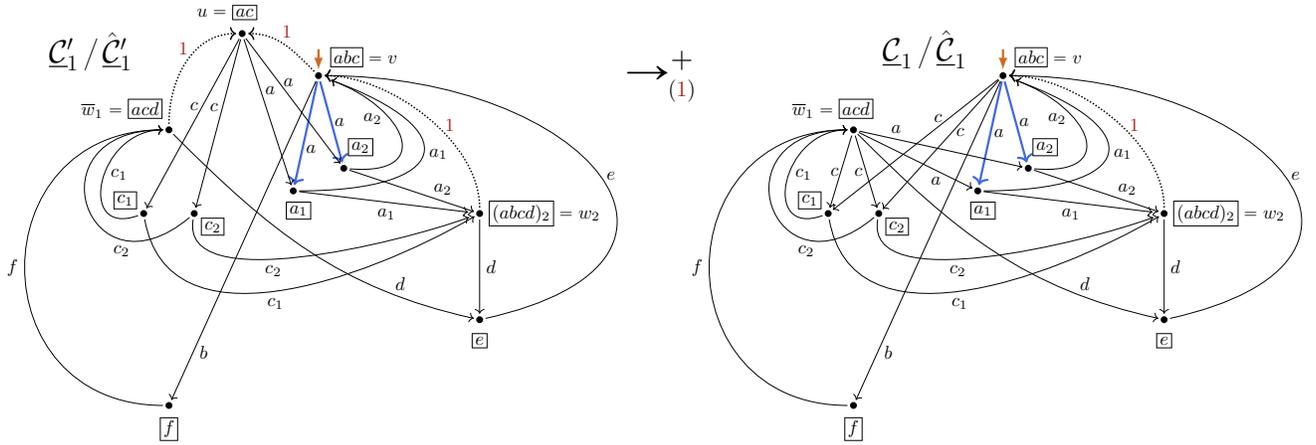

But we note now that no loop-entry transition is created at $u$ in $\underline{\mathcal{C}}'_1$, and so the added 1-transitions are not back-links. Therefore this refinement step is not one into the direction of the (assumed) 1-transition limited LLEE-witness $\hat{\underline{\mathcal{D}}}$.

It would be possible, however, to share only a non-empty subset of the four transitions at $u$. For instance, when sharing only the $c$-transitions, the 1-chart $\underline{\mathcal{C}}'_1$ below on the left is obtained:

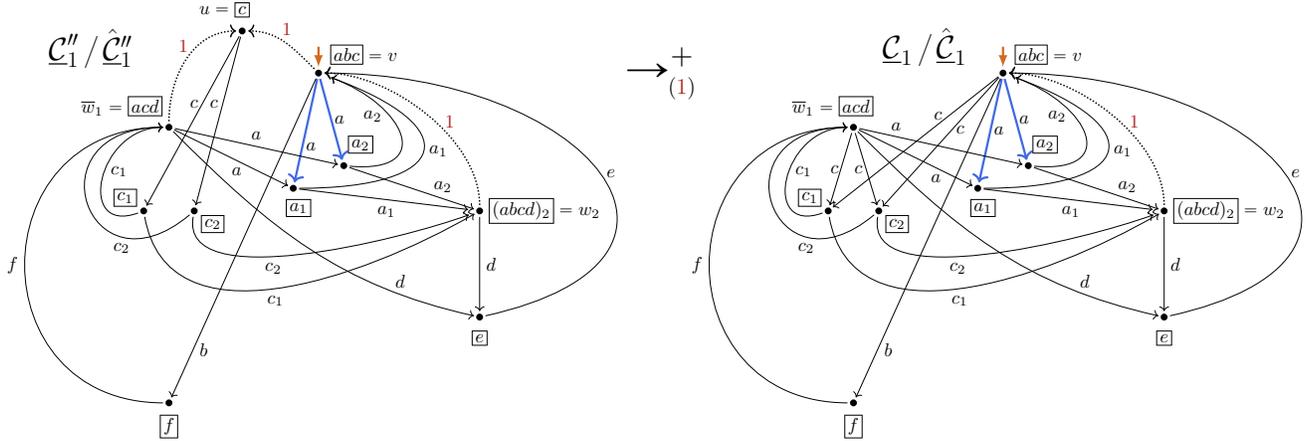

But also now there is not a loop-entry transition arising at $u$, either.

It is clear from these two examples that also for all non-empty subsets of $a$- and $c$-transitions from $v$ and $w_1$ the operation of sharing them via the new vertex $u$ does not lead to a loop-entry transition at $u$: maximal paths from $u$ must, in order to be able to return to $u$, reach $\overline{w}_1$ or $v$ first; but there they can always avoid taking one of the 1-transitions back to $u$. Therefore none of these options can lead to a 1-transition refinement in which the introduced 1-transitions are back-links, as they would have to be in order to be part of the 1-transition limited LLEE-witness $\hat{\underline{\mathcal{D}}}$ of the 1-chart $\underline{\mathcal{D}}$ that we assumed refines $\underline{\mathcal{C}}_1$.

At this point we must recognize that it is not possible to add such a 1-transition in order to refine $\underline{\mathcal{C}}_1$ further towards the LLEE-1-chart $\underline{\mathcal{D}}$ with LLEE–1-lim, as assumed. Our assumption that this is possible cannot be upheld.

Thus we have shown the part of (S2) now also for $\underline{\mathcal{C}}_1$. □

**Example** (*Combining the refinements of $\mathcal{C}_{10}$ into $\underline{\mathcal{C}}_1$ and into $\underline{\mathcal{C}}_2$ does not yield a loop sub-1-chart, let alone a* LLEE*-1-chart*). In the proof of Lem. 5.6 we dismissed the option of a refinement that combines those refinement steps that lead from $\mathcal{C}_{10}$ in Ex. 5.3 to the 1-charts $\underline{\mathcal{C}}_1$ in Ex. 5.3 and to $\underline{\mathcal{C}}_2$ in the proof of that lemma. But it can be instructive to see concretely why we could do that, and reassuring to recognize that, apart from not leading to a 1-chart with LLEE–1-lim, there is not even a single loop sub-1-chart created.

By adding to $\mathcal{C}_{10}$ the 1-transition in $\underline{\mathcal{C}}_1$ as well as the 1-transition in $\underline{\mathcal{C}}_2$, and by removing the transitions that have been eliminated from $\mathcal{C}_{10}$ to $\underline{\mathcal{C}}_1$ and from $\mathcal{C}_{10}$ to $\underline{\mathcal{C}}_2$ (the $a$- and $c$-transitions from $w_2$ in both of these steps, the $b$-transition from $w_2$ in the first step, and the $d$-transition from $w_2$ in the second step) we obtain the 1-chart $\underline{\mathcal{C}}_{1/2}$ of the form below left:

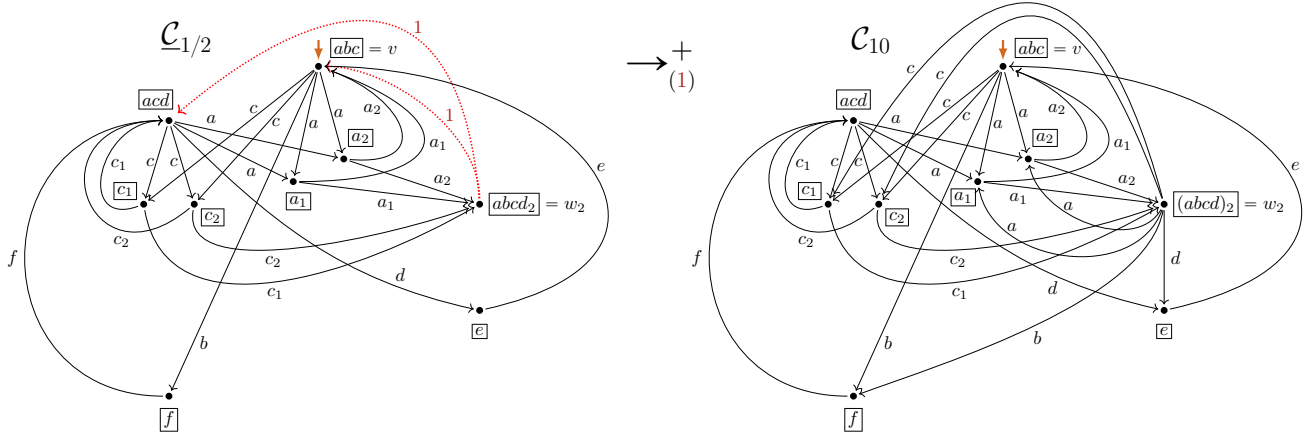

The 1-chart $\underline{\mathcal{C}}_{1/2}$ is obviously a refinement of $\mathcal{C}_{10}$, which arises by from $\underline{\mathcal{C}}_{1/2}$ by two $\rightarrow_{(1)}$ steps. Furthermore, every vertex of $\underline{\mathcal{C}}_{1/2}$ is a proper-transition target.

But $\underline{\mathcal{C}}_{1/2}$ is *not* a 1-chart with LLEE–1-lim, because (as stated in the proof of Lem. 5.6): if there were a 1-transition limited LLEE-witness of $\underline{\mathcal{C}}_{1/2}$, then the vertex $w_2$ would have back-links to two different loop vertices, which is not possible in a LLEE-witness. For this reason $\underline{\mathcal{C}}_{1/2}$ cannot be refined into a 1-chart with LLEE–1-lim, either.

What is more, $\underline{\mathcal{C}}_{1/2}$ is not a LLEE-1-chart at all, because (in view of Prop. 4.4) it does not satisfy LEE. To verify this, we check that none of the transitions of $\underline{\mathcal{C}}_{1/2}$ is a loop-entry transition. Indeed, for every transition $\tau = \langle u, \boldsymbol{a}, u' \rangle$ in $\underline{\mathcal{C}}_{1/2}$ there is an infinite path from $u'$ that does not visit the source $u$ of $\tau$. Hence no transition induces a loop subchart. Therefore $\underline{\mathcal{C}}_{1/2}$ does not possess loop subcharts. But as $\underline{\mathcal{C}}_{1/2}$ exhibits infinite behavior, it follows that $\underline{\mathcal{C}}_{1/2}$ does not satisfy LEE.

Therefore via the combination of refinements from $\mathcal{C}_{10}$ to $\underline{\mathcal{C}}_1$ and to $\underline{\mathcal{C}}_2$ not only no progress has been made towards a refinement of $\mathcal{C}_{10}$ with LLEE–1-lim, but also no progress towards a refinement of $\mathcal{C}_{10}$ into a LLEE-1-chart.